\documentclass[11pt]{article}
\usepackage{authblk} 
\usepackage{times}
\usepackage{longtable}
\usepackage{graphicx}
\usepackage{array}
\usepackage{amssymb}
\usepackage{listings}
\usepackage{soul}
\usepackage{tikz}
\usepackage{amsmath}
\usepackage{mathtools}
\usepackage{dsfont}
\usepackage{relsize}
\usepackage{yfonts}
\usepackage[T1]{fontenc}
\usepackage{graphics}
\usepackage{booktabs}
\usepackage{color}
\usepackage{lipsum}
\newcommand\myeq{\mathrel{\stackrel{\makebox[0pt]{\mbox{\normalfont\tiny def}}}{=}}}
\DeclareMathAlphabet\mathbfcal{OMS}{cmsy}{b}{n}
\newcommand{\Conv}{\mathop{\scalebox{1.16}{\raisebox{-0.2ex}{$\ast$}}}}%

\begin{document}

\title{A New Conceptual Framework for the Therapy 
\\      by Optimized Multidimensional Pulses of Therapeutic Activity. 
\\     The case of Multiple 
Myeloma Model.}  

\author{Denis Horvath\\
Technology and Innovation Park, Centre of Interdisciplinary Biosciences,\\ 
Faculty of Science, P.~J.~\v{S}af\'arik University,  \\ 
Jesenn\'a 5, 04154 Ko\v{s}ice, Slovak Republic  \\
horvath.denis@gmail.com 
\and
Branislav Brutovsky\\
Department of Biophysics, Faculty of Science, 
P.~J.~\v{S}af\'arik University, \\ 
Jesenn\'a 5, 04154 Ko\v{s}ice, Slovak Republic \\ 
branislav.brutovsky@upjs.sk}

\maketitle

\begin{abstract}
We developed simulation methodology to assess eventual therapeutic
efficiency of exogenous multiparametric changes in a four-component
cellular system described by the system of ordinary differential 
equations.  The method is numerically implemented to simulate the
temporal behavior of a cellular system of multiple myeloma cells.
The problem is conceived as an inverse optimization task where the
alternative temporal changes of selected parameters of the ordinary
differential equations represent candidate solutions and the objective
function quantifies the goals of the therapy. The system under study  
consists of two main cellular components, tumor cells and 
their cellular environment, respectively. The subset of model parameters 
closely related to the environment is substituted by exogenous
time dependencies - therapeutic pulses combining continuous
functions and discrete parameters subordinated thereafter
to the optimization. Synergistic interaction of temporal parametric 
changes has been observed and quantified 
whereby two or more dynamic parameters show effects that absent 
if either parameter is stimulated alone.
We expect that the theoretical insight into unstable tumor growth 
provided by the sensitivity and optimization studies could, eventually,
help in designing combination therapies.
\end{abstract}


\vspace*{3mm}

\section{Introduction}

Mathematical models of cancer are widely used to get insight into dynamics
of cancer initiation and progression whereby they help researchers to design
new therapeutical strategies \cite{Preziosi2003,Gatenby2003b,Roose2007,Bellomo2008}.
As a consequence of ongoing accumulation of relevant biological knowledge,
as well as availability of prognostic variables, complexity of mathematical models 
in tumor biology constantly increases. At the same time, the bottom-up derivation 
of the relation between microscopic level and macroscopic 
behavior is far from straightforward.
The balance between biological relevance and mathematical complexity
is typically achieved by means of iterations containing
addition or omission of algebraic expressions in the respective model.
Despite the vast majority of iterative model-modifying operations is motivated by
relatively well understood short-term effects of the respective 
variables and algebraic terms, too many details can, paradoxically,
make the overall interpretation of the 
numerical results more difficult \cite{Transtrum2014}.  

Due to nonlinear interactions, that pose the most difficult obstacle
(except the fixed points), the long-term or large-scale 
outcomes cannot be usually interpreted without detailed simulations.
Along with progress in oncological experimental studies and synthesizing of available information, emerging field of multi-scale modeling \cite{Sanga2006} provides novel 
computational strategies of exploring cancer biology simultaneously with anticancer therapies.
Although this type of approach enables to reduce the number of degrees of freedom, 
the problem associated with too many parameters and their relevance
persists, which is the common problem in most
cancer simulation models and methodologies, thus requiring to address 
the issue  of parametric uncertainty,  unidentifiability 
and relevance \cite{Transtrum2014,Raman2017}. 

Motivated by these general unresolved questions of parametric importance, we developed 
the specific theoretical approach based on the model
of bone remodeling by Komarova et al. \cite{Komarova2003} and its phenomenological
extension applied to multiple myeloma (MM) bone disease \cite{Koenders2016}.
The above models illustrate the association of incremental incorporation
of novel biological information with an increase in the number of the 
parameters.  It follows that to control particular scenario (outcome) 
by the parameters of the model, it is essential
to estimate the relevance of the specific parameters and 
their meaningful groupings. The issue of parametric relevance is especially challenging
from the point of view of the therapy design. Within the context of mathematical models
of cancer or tumor growth, there is a systematic effort 
towards identification of the appropriate parameters in order
to achieve, hopefully, therapeutic effects by
parametric modifications \cite{Preziosi2003}.

The leitmotiv of here proposed methodology 
consists in the idea that coordinated multiparametric (high-dimensional) changes can positively, in a therapeutic
sense, effect cell populations.
Most of the previous theoretical studies of cancer 
have focused primarily on the virtual therapeutic interventions performed via 
variations of parameters which describe drug-induced 
proliferation, necrosis or apoptosis (see Ref.\cite{Preziosi2003} 
and references therein). Since non-linear systems, including the cancer models \cite{Yafia2006}, 
show a wide range of emerging and unstable behaviors, 
we assume that not yet fully understood area of therapeutic options can be much more fruitful
and more structured than intuitively expected \cite{Gatenby2003V}.

To overcome the existing barriers to treatment, 
such as the resistance to chemotherapy, it seems important to explore
more advanced and universal strategic options 
(see e.g. \cite{Gatenby2009a,Gatenby2009b,Gillies2012}).
Deeper understanding of their impact might be achieved by
combining knowledge from complementary 
research fields, such as optimization theory, inverse problems 
\cite{Engl2009}, and sensitivity analysis 
\cite{Gutenkunst2007,Ingalls2008,Marino2008,Sahle2008,Zi2011,Babtie2014,Zhang2015,Ferretti2016,Borgonovo2016}.

In the present study we pursue the idea of multidimensional investigation
of the global parametric sensitivity of cancer population models.
Our method extends the range of methods 
for the sensitivity analysis of the systems of nonlinear ordinary differential 
equations (ODEs).  Therefore, instead of application of standard 
analysis of the respective ODEs system, 
we are interested in the specific global parametric sensitivity analysis \cite{Ferretti2016} 
which can be seen as a way to propose 
a qualitatively new therapies. Despite obvious 
differences in the technical and implementation details,
simultaneous use of the optimization and sensitivity analysis 
is commonly applied feature \cite{Gutenkunst2007,Sahle2008}.

To put forward the above ideas, we apply them to analyze MM bone
disease. In healthy bone tissue, the bone homeostasis is maintained
(or re-established after fracture or microscale damages) by many coordinated
actions at cellular and molecular levels, summarily called the bone 
remodeling \cite{Hadjidakis2006}. The process maintains the balance between
removal of the bone tissue (resorption) and the formation of new tissue (ossification).
The bone remodeling involves two crucial cell types - osteoclasts (OCs) and
osteoblasts (OBs). The former cells are responsible for disassembling the old bone
tissue, the latter for synthesizing the new tissue.
The imbalance between intensities of the two processes 
leads to bone diseases, such as osteoporosis or bone cancer.
Physiological numbers of OCs and OBs are guaranteed by their mutual influence 
via autocrine and paracrine signaling. Due to serious health implications of the 
MM cancer, several mathematical models were developed to understand 
nonlinear dynamics of the OCs and OBs production, homeostasis  
and decay \cite{Dingli2009,Ryser2009,Ayati2010,Qiao2015,Bouchnita2016}, 
proposing alternative equations to describe the system.
The assembly composed of many coupled submodels \cite{Peterson2010} 
has been considered to adequately describe calcium homeostasis 
and intracellular signaling as occurring in the process of bone remodeling.  
The analysis of homeostatic control also means a better understanding of MM, 
which is often manifested by hypercalcemia \cite{Zomas2014}. 

Here presented computational considerations derive from the
ODEs model by Komarova et al. \cite{Komarova2003}
describing population dynamics of OCs and OBs in healthy tissue. 
Their parametrization of the net effectiveness of the 
OCs and OBs autocrine and paracrine factors, such as 
TGF$-\beta $ (one of the classes of polypeptide transforming  
growth factors), RANKL (the receptor activator of nuclear 
factor kappa-B ligand), and OPG (osteoclastogenesis inhibitory factor),
was used in the recent mathematical model of multiple myeloma 
by Koenders and Saso \cite{Koenders2016}. 
Therein, the authors augmented the above model by Komarova et al. \cite{Komarova2003} 
by including the population of myeloma cells with the
feedback to the OCs proliferation and the OBs decay into the
model. Moreover, the population of so-called 
joint cells (JCs) also called OCs-MM 
hybrid cells \cite{Kupisiewicz2011,Andersen2007}, 
formed when one OC and one MM cell meet, 
was included into the extended set of equations 
\cite{Koenders2016}. This model 
also includes the tumoral feedback on the paracrine 
interactions, interactions of OBs and OCs \cite{Lemaire2004}. 
The OCs-MM hybrid cells co-cultures are considered to contribute significantly 
to the formation of bone-resorbing OCs and bone destruction in the MM case. 
Although widely discussed phenomena of intratumoral MM heterogeneity 
\cite{Bouchnita2016,Tang2016}, phenotypic plasticity \cite{Qiao2015} 
and drug resistance can decisively influence the treatment, we leave 
these aspects to further research. On the other hand, formation of JCs belonging to the wide class 
of pathological cell fusion processes in cancer 
\cite{Nuobissi2016,Searles2018}, already captured
at some basic level in \cite{Koenders2016},
can be considered as an alternative mechanism contributing 
to the phenotypic heterogeneity thereby increasing the chemical 
resistance and metastatic potential. 
Promising approach to the MM analysis represents the agent based model introduced in \cite{Qiao2015} 
which incorporates DKK1-Wnt-OPG/RANKL pathway and cytokine stimulation.  The main strength of 
the model consists in the study of therapeutic effects of Lidamycin, glucocorticoids and
Anti-DKK1 mAb (BHQ880). Unfortunately, it is rather difficult to compare or associate 
this type of models to ODEs. 

In the present work we address the problems discussed in 
\cite{Koenders2016} from different perspectives and research interests. The emphasis 
is placed on the specific aspects of the sensitivity analysis, parametric relevance, 
and optimization of therapeutic interventions. Here applied 
variant of the sensitivity analysis uses therapeutically relevant 
parametric moves (pulses) determined as the worst-case optimization outputs.

To develop the efficient therapeutic schedule, one implicitly faces to several, 
sometimes conflicting \cite{Costa97} or overlapping, objectives which must be 
considered simultaneously.  Regarding this, our motivation for the optimization 
of parametric combinations is, apart from the search for the efficient solution, to formulate 
the methodological multi-objective framework that may provide basis for further studies. 
Despite intuitive plausibility of the multi-objective formulation, solutions of the 
majority of multi-objective problems are not straightforward, as different paradigms 
often lead to different solutions. There is a variety of methods how to quantify the quality 
of the respective solution which is expected to fulfill simultaneously a few objectives 
\cite{Miettinen2002}. Among them, the scalarization approach \cite{Ehrgott2006}
which leads to a compromise single-objective formulation is one of the widely 
used practices. From the point of view of the reliability of the results, the degree of consistency (stability) 
achieved through diverse scalarizations is important. In the most favorable circumstances, 
the scalarization points to the single solution on the Pareto front. 

In our paper, in addition to detailed analysis of the worst-case "minimax" scalarization, 
we present the alternative numerical experiments 
with the optimization of single-objective functions 
based on the aggregation transforms of several objectives \cite{Beliakov2015}.
In order to avoid uncertainty in the interpretation of our results, 
we use optimization that does not use stochastic sampling and is
intentionally limited to generating deterministic outputs.
To analyze the respective non-differentiable scalarizations, 
we employed the comprehensive direct grid search which evaluates the 
pair  of purposefully constructed objective 
functions at each grid point of some feasible parametric region. 
This choice reflects the recent trends in 
machine learning \cite{Claesen2015,Wistuba2015}
where the grid search variants also known 
as {\em hyperparameter searches} receive increasing attention. 

The main aim  of our study is to bring
interesting alternative approach 
which could stimulate further research in the respective direction 
instead of demonstrating superiority of the specific optimization algorithm.
Systematic analysis performed by the grid search technique can help to find parametric
boundaries within which improvements can be done (even if highly diluted grids are used). 
We note that, at this stage, the independence of the grid search
technique on the initial conditions with no tendency of being
trapped in a local optimum vindicates its (low) computational performance 
(nevertheless, in sec.~\ref{sec:manifOpt} we discuss the hybrid 
discrete-continuous optimization methodology which has significant potential 
to improve the coarse grid results).

The theoretical background to our considerations 
is the four cell population model \cite{Koenders2016} described by the system of four ODEs written 
in the normal form  
\begin{eqnarray} 
\frac{d C}{dt} &=& R_C(T,C,B)\,,
\qquad 
\frac{d B}{dt}=  R_B(T,C,B)\,, 
\label{eq:CBTJ}
\\
\frac{d T}{dt} &=& R_T(C,T,J)\,,\qquad 
\frac{d J}{dt}=  R_J(C,T,J)\,, 
\nonumber
\end{eqnarray}
where the respective rates of change of the four populations,
$C, B, T, J$, referring to the population of OCs ($C$), 
OBs ($B$), MM ($T$) and JC ($J$), 
respectively, have the explicit form 
\begin{eqnarray}
R_C 
&\equiv& \alpha_C (1+ h_{CT} T \,) C^{g_{CC}} B^{g_{CB}} -  \beta_C C - \alpha_J C T  \,, 
\label{eq:RCRBRTRJ}
\\
R_B 
&=&  
\alpha_B (1- h_{BT} T\,) C^{g_{BC}} B^{g_{BB}} - \beta_B B\,,
\nonumber
\\
R_T  
&\equiv&
\alpha_T C^{g_{TC}} T^{g_{TT}} - \beta_T T - \alpha_J C T + \kappa_J J
\,,
\nonumber 
\\
R_J &\equiv & \alpha_J C T - \beta_J J\,.
\nonumber
\end{eqnarray}
This autonomous system couples population number 
of the OCs, MM cells, and JCs in the representative 
volume of bone marrow basic multicellular units (BMU);
$\alpha_{\bullet}$,$ \beta_{\bullet}$ 
represent activities of the cell production and 
removal. The power-law nonlinear structure proposed by \cite{Komarova2003}
of the interactions is parametrized by $g_{\bullet \bullet}$ which represent 
the effectiveness of the OCs and OBs autocrine and paracrine factors. 
Increased sensitivity of OCs sand OBs cells due to the influence of MM cells
is modeled by $h_{CT}$ parameter.
The interpretation behind the term $h_{CT} T C^{g_{CC}} B^{g_{CB}} $ is that MM cells 
stimulate bone metabolism and bone marrow micro-environment 
by means of RANKL, decrease in OPG expression and production of chemokines
MIP$-1\alpha$ (human macrophage inflammatory protein), 
MIP$-1\beta$ (macrophage inflammatory protein-1),
and SDF$-1\alpha $  (stromal cell-derived factor) 
\cite{Koenders2016,Heider2005}. Furthermore, the term 
$ (- h_{BT}) T C^{g_{BC}}  B^{g_{BB}} $
describes how MM cells suppress OB function by the secretion of OB inhibiting factors, 
such as Wnt inhibitors DKK$-1$ and sFRP$-2$. 
The parameter $\kappa_J$ in the term $\kappa_J J$ denotes the backward transformation
rate of the JCs into MM cells with the specific assumption \cite{Koenders2016} that while MM 
cells survive dissociation, OCs are not recreated, which means that 
there is no adequate term $\propto J$ in the $R_C$ \cite{Koenders2016}.

The paper is organized as follows. In the section \ref{sec:methodology}
we introduce basics of the methodology including the parametrization
of symmetric pulses and the form of the objective function.
In the section \ref{sec:num} we present numerical results obtained for given
parametric settings.  The alternative scalarizations of the
multi-objective problem are discussed in sec.\ref{sec:alters}. 
In section \ref{se:lamPen} the alternative objective function with  the regularization 
penalty term to minimize toxic side effects (at some
stylized level) of the therapies is studied. The model of the more
realistic exogenous asymmetric therapeutic pulses is introduced
in sec.~\ref{sec:APulse}, where we also provide extensive
comparison of the impact of symmetric and asymmetric pulse forms, respectively,
on the optimum obtained. In addition, there are presented results of the optimization performed on the
selected manifolds (which go beyond the results provided by grid
search). The generalization of synergistic quantification of the
parametric pairs is presented as well.  Finally, the conclusions are
presented. Two appendices provide additional information about the
robustness of optima in periodic environments and numerical accuracy
of the calculations.

\section{The methodology}\label{sec:methodology}

In this section we provide methodological details of our approach.
The main aim of our work is to formalize and analyze the time-varying influence
of the populations of OCs and OBs (viewed as the environment)
on the populations of tumoral cells (including the joint cells)
via autocrine and paracrine interactions and, consequently, to exploit
this framework to drive dynamics of the tumor cells population
towards required direction.

\subsection{The equilibrium approximation and short-time dynamics}

Without sufficient knowledge of symmetry or invariance, or without
significant simplifications
the numerical solution is, in principle, the only universal option
to study population dynamics (Eqs.~\ref{eq:CBTJ}) under very general nonlinear 
conditions. Nevertheless, some partial insights into the tumor behavior can be achieved
without actually solving the corresponding ODEs numerically 
by starting with the static equilibrium analysis and then proceeding with 
approximate dynamic considerations.

The information about equilibrium enables to derive long-term trends in overall
dynamics. Although the system of the transcendental equations
$ R_{X} = 0 $; $X \in \{C, B, T, J \} $ 
with $R_X $ given by Eq.(\ref{eq:RCRBRTRJ})
presents no computational difficulty, 
the problem is that the system does not provide 
steady-state solutions (fixed points) in the explicit form.
Analytical solvability can be achieved exclusively
for the choice $g_{TC}=0$ and $J=0$ \cite{Ayati2010,Koenders2016}.
Discussion about the stability is postponed to the subsection 
\ref{sec:eqstab}. Despite its biological 
limitations, the above approximation is 
useful as an initial guess for more advanced formula.
In such case the equilibrium solution may be expressed as 
\begin{eqnarray} 
T^{(0)}_{eq} 
&=&
\left(
\frac{\alpha_T}{\beta_T} 
\right)^{\frac{1}{1-g_{TT}}}\,,
\label{eq:scalform1}
\\
C^{(0)}_{eq} 
&=&
{\mathcal{G}}\left(\frac{\alpha_C}{\beta_C},\frac{\alpha_B}{\beta_B},
\frac{\alpha_T}{\beta_T},1-g_{BB},g_{CB},g_{xy}\right)\,,
\nonumber
\\
B^{(0)}_{eq}  &=&
{\mathcal{G}}\left(\frac{\alpha_C}{\beta_C},\frac{\alpha_B}{\beta_B},
\frac{\alpha_T}{\beta_T},g_{BC},1-g_{CC},g_{xy}\right)\,,
\nonumber 
\end{eqnarray}
where
\begin{eqnarray}
{\mathcal{G}}(x,y,g_x, g_y, g_{xy})
&=& [ x (1 + h_{CT} T^{(0)}_{eq} )]^{\frac{g_x}{g_{xy}}}
+  [ y (1 - h_{BT} T^{(0)}_{eq} )]^{\frac{g_y}{g_{xy}}}
\label{eq:scalform2}
\end{eqnarray}
is the auxiliary function 
with the fixed argument $g_{xy} \equiv g_{CC} ( g_{BB} -1 ) -  g_{CB} g_{BC} +1 $. 
The formula emphasizes the scaling dependence on the ratios of the proliferative/apoptotic rates 
$\alpha_{T}/\beta_{T}$,  $\alpha_{B}/\beta_{B}$,  $\alpha_{C}/\beta_{C}$
and the exponents $(1-g_{BB})/g_{xy}, $ $g_{CB}/g_{xy}, $ $g_{BC}/g_{xy},$ $(1-g_{CC})/g_{xy}$. 
The equilibrium becomes, obviously, not tumor-free, as $\alpha_T\neq 0$. 
When the simplifying assumption $g_{TC} = 0$ is relaxed,
the reliability of the formula for $T^{(0)}_{eq}$ can be further improved 
in an iterative way by revisiting condition $R_T=0$ and using 
the initial approximation $C\simeq C^{(0)}_{eq}$. 
Owing to the above steps, the improved approximation 
for tumoral population number 
$T^{(0,+)}_{eq}$ reflects the
impact of OCs as follows 
\begin{eqnarray}
T^{(0,+)}_{eq}  &=& T^{(0)}_{eq} \left[
 \frac{(C^{(0)}_{eq})^{g_{TC}}}{
1+\frac{\alpha_J}{\beta_T} C^{(0)}_{eq} 
(1-\frac{\kappa_J}{\beta_J})}\,\right]^{\frac{1}{1-g_{TT}}}\,,
\\
&&
\nonumber 
\\
J^{(0,+)}_{eq} &=& \frac{\alpha_J}{\beta_J}\, C^{(0)}_{eq} T^{(0,+)}_{eq}\,, 
\nonumber
\end{eqnarray}
where the population number $J_{eq}^{(0,+)}$ is obtained from $R_J=0$.
In addition, we see that the above improvements of Eq.(\ref{eq:scalform1})
exhibit the scaling form with arguments 
$\alpha_J/\beta_T$, $\kappa_J/\beta_J$ and $\alpha_J/\beta_J$. 
The approximate equilibrium solution of this type 
implicates the possibility of exploiting the influence 
of environmental characteristics $C^{(0)}_{eq}$  
to affect the population of tumoral 
cells described by the $T^{(0,+)}_{eq}$ and $J^{(0,+)}_{eq}$ variables. 

The essence of the environmental 
concept can also be captured analytically 
by examining a short-time nonequilibrium picture of environmental influence, 
which is discussed below.
The straightforward quantification of the endogenous ($edg$) tumor response
to the environment can be quantified by the two-time population ratios defined by
\begin{eqnarray}
f^{edg}_Y(t,t+\Delta t) \myeq
\frac{Y(t+\Delta t)}{Y(t)}\,,\qquad Y(t )\in \{T(t),J(t)\}\,
\end{eqnarray}
considered for both tumoral population 
variants $Y(t)$ and sufficiently small $\Delta t$. 
In subsection \ref{sec:tumrespons} is the above preliminary 
concept revisited [see Eq.(\ref{eq:fTJ}) in further]
and discussed for larger separation intervals, i.e. for $\Delta t\rightarrow (t_E-t_S)$.

Deeper understanding of the linkages between 
the environmental populations $B, C$ 
and the responses of population $T$ 
can be obtained simply by using the truncated 
short time Taylor expansion 
\begin{eqnarray}
f^{edg}_Y(t,t+\Delta t) = 
1 +  \frac{\Delta t \,R_Y(t)}{Y(t)} +
\frac{(\Delta t)^2}{2 Y(t)}\frac{d R_Y}{dt} + 
\mathcal{O}(\Delta t^3)\,.
\label{eq:fedg}
\end{eqnarray}
As will become clear later on, sufficient evidence 
of environmental influence 
can be obtained by expanding 
$f^{edg}_Y(t,t+\Delta t)$ into $\Delta t^2$ order at least.  

The first order represented by $dT/dt=f_T^{edg} \sim R_T(C,T,J)/T$, $dJ/dt=f_J^{edg}\sim R_J(C,T,J)/J$
confirms that $C$ influences $dT/dt$ and $dJ/dt$ 
while the impact of the changes due to 
environmental variable $B$ remains hidden. 
By means of the straightforward differentiation, the 
coefficients $\sim \frac{d R_T}{dt}$, $\frac{d R_J}{dt}$
corresponding to $\Delta t^2$ order from Eq.(\ref{eq:fedg}) 
can be expressed 
\begin{eqnarray}
\frac{1}{T}\frac{d R_T}{dt} 
&=& 
\alpha_T C^{g_{TC}-1}  T^{g_{TT}-2}  (  g_{TC} T R_C + g_{TT} C R_T  )
\\ 
&-& \alpha_J \left( 
R_C + \frac{C}{T} R_T \right)  +   \kappa_J \frac{R_J}{T} \,, \nonumber
\\
\frac{1}{J}\frac{d R_J}{dt}  
&=& 
\alpha_J \left( \frac{T}{J} R_C  +  \frac{C}{J}  R_T\right)  -  \beta_J \frac{R_J}{J}\,.
\nonumber 
\end{eqnarray}
As both the above right-hand sides 
include $R_C\equiv R_C(T,C,B)$,  
the respective rates 
depend not only on $C$ 
(as in the first order case) 
but on $B$ as well, which makes the causal relation between 
environmental $(C,B)$ and tumoral ($T, J$) populations more explicit. 
The short-time causality between exogenous 
parameter-induced environmental changes
and tumor subsystems that closely 
applies to the present work 
can be captured in an analogous way.
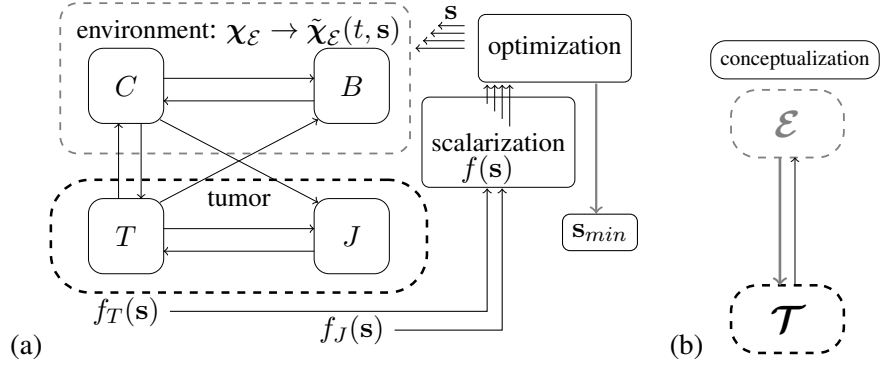
\begin{figure}[bth]
\begin{center}
\begin{tabular}{cc}
(a)
\begin{tikzpicture}
\node[draw,rectangle, rounded corners=0.2cm, minimum size=1cm, line width=0.2] at (0,0){$T$}; 
\node[draw,rectangle, rounded corners=0.2cm, minimum size=1cm, line width=0.2] at (3,0){$J$}; 
\node[draw,rectangle, rounded corners=0.2cm, minimum size=1cm, line width=0.1] at (0,2){$C$}; 
\node[draw,rectangle, rounded corners=0.2cm, minimum size=1cm, line width=0.1] at (3,2){$B$}; 
\node at (1.5,2.73) {{\small environment}: $\boldsymbol{\chi}_{\mathcal E} \rightarrow $
$\tilde{\boldsymbol{\chi}}_{\mathcal E} (t,{\bf s})$ }; 
\node[draw,dashed,rectangle, rounded corners=0.23cm, minimum width=4.65cm, minimum height=2cm, line width=0.24mm, gray] at (1.45,2.1){}; 
\node at (1.5,0.5) {\small tumor}; 
\node[draw,dashed,rectangle, rounded corners=0.52cm, minimum width=5cm, minimum height=1.5cm, line width=0.34mm] at (1.5,-0.0){}; 
\draw[thick,->,line width=0.1mm] (-0.1,0.5) -- (-0.1,1.5); 
\draw[thick,->,line width=0.1mm] (2.5,1.8) -- (0.5,1.8); 
\draw[thick,->,line width=0.1mm] (0.5,2.1) -- (2.5,2.1);
\draw[thick,->,line width=0.1mm] (0.45,0.45) -- (2.55,1.55); 
\draw[thick,->,line width=0.1mm] (0.2,1.5) -- (0.2,0.5); 
\draw[thick,->,line width=0.1mm] (2.5,-0.2) -- (0.5,-0.2); 
\draw[thick,->,line width=0.1mm] (0.5,0.1) -- (2.5,0.1);
\draw[thick,->,line width=0.1mm] (0.45,1.55) -- (2.55,0.45); 
\node (fJ) at (3, -1.25) {$f_J({\bf s})$};
\node (fT) at (0, -1.0) {$f_T({\bf s})$};
\node (tarfJ)  at (5.0,0.77){};   
\node (tarfT)  at (4.8,0.77){};   
\node[draw,rectangle, rounded corners=0.1cm, line width =0.05mm, minimum width=1.5cm, minimum height=1.2cm] at (4.95,1.25){\small scalarization}; 
\node (scf2)  at (4.8,0.89){$f({\bf s})$};
\draw[->, to path={-|(\tikztotarget)}] (fT) edge (tarfT) ;
\draw[->, to path={-|(\tikztotarget)}] (fJ) edge (tarfJ) ;
\node[draw,rectangle, rounded corners=0.1cm, line width=0.05 mm, minimum width=1.5cm, minimum height=0.9cm] at (5.7,2.5) {\small optimization}; 
\node (fJ) at (4.32, 2.98) {${\bf s}$};
\draw[thick,->,line width=0.07mm] (4.5,2.8) -- (4.15,2.8); 
\draw[thick,->,line width=0.07mm] (4.5,2.7) -- (4.05,2.7); 
\draw[thick,->,line width=0.07mm] (4.5,2.6) -- (3.95,2.6); 
\draw[thick,->,line width=0.07mm] (4.5,2.5) -- (3.85,2.5); 
\draw[thick,->,line width=0.04mm] (4.8,1.76) -- (4.8,2.01); 
\draw[thick,->,line width=0.04mm] (4.9,1.66) -- (4.9,2.01); 
\draw[thick,->,line width=0.04mm] (5.0,1.56) -- (5.0,2.01);
\draw[thick,->,line width=0.04mm] (5.1,1.46) -- (5.1,2.01); 
\draw[thick,->,line width=0.3mm, gray] (6.25,2.03) -- (6.25,0.3); 
\node[draw,rectangle, rounded corners=0.1cm, minimum width=0.5cm, minimum height=0.4cm] at (6.3,0.05){${\bf s}_{min}$}; 
\end{tikzpicture}
& (b)
\begin{tikzpicture}
\node[draw,rectangle, dashed, gray, rounded corners=0.35cm, minimum width=1.5cm, minimum height=0.9cm, line width=0.24mm] at (0,2.9){\Large $\mathbfcal{E}$}; 
\node[draw,rectangle, dashed, rounded corners=0.35cm, minimum width=1.5cm, minimum height=0.9cm, line width=0.34mm] at (0,0.3){\Large $\mathbfcal{T}$};
\draw[thick,->,line width=0.35mm, gray] (-0.1,2.45) -- (-0.1,0.75); 
\draw[thick,->,line width=0.1mm] (0.1,0.75) -- (0.1,2.45);
\node[draw,rectangle, rounded corners=0.2cm, minimum width=1.5cm, minimum height=0.3cm, line width=0.02mm] at (0.1,3.75){\scriptsize conceptualization}; 
\end{tikzpicture}
\end{tabular}
\caption{Schematic plot showing 
interactions (oriented graph edges) between the population numbers $C,B,T,J$, 
as described by Eqs.(\ref{eq:CBTJ}), (\ref{eq:RCRBRTRJ}).  
The topological representation of the algebraic 
structure of interactions and direct causal relations 
is presented in the part (a). As $R_T(C,T,J)$ 
does not include $B$, the directed link 
between nodes $B$ and $C$ absents.
The links belong to the dependencies of 
the population rates 
[see Eqs.(\ref{eq:CBTJ})]. The dynamics of environment is affected by the six pivotal 
parameters from Eq.(\ref{eq:listpar}) 
later replaced by smooth exogenous pulses
[see Eqs.(\ref{eq:partim}),(\ref{eq:tildhCT})]. The scheme also clarifies the influence 
of discrete ${\bf s}$ (string) parameters from 
Eq.(\ref{eq:svec})
which determine the relations between 
$C, B$ (environment) and the tumoral populations  
$T$, $J$. The scheme is supplemented by the 
computational aspects related to the choice 
of the objective functions $f_T, f_J$, their 
scalarization (see subsection \ref{sec:tumrespons} and sec.\ref{sec:alters}) and optimization 
leading to the optimal ${\bf s}_{min}$ [see Eq.(\ref{eq:smin})]. 
In part (b) we depict the nodes corresponding to the 
conceptual framework from section \ref{sec:Conc} 
where the only pair of abstract population 
vectors $\mathbfcal{T}(t)$, 
$\mathbfcal{E}(t)$  admits encompassing more general domain
of population models with environments.}
\label{fig:ff1}
\end{center}
\end{figure}

\subsection{Sensitivity analysis and inverse/optimization task of ODEs system}\label{sub:sensit}

The utmost ambition of our work is the proposal of a numerical procedure that can be viewed as a kind of global sensitivity analysis 
designed for specific ODEs. 
We identify the most sensitive parameters or their combinations 
as potential intermediaries for an indirect 
virtual therapeutic intervention. For our purposes, 
which are mainly illustrative, the time-consuming search for the most influential and,
hopefully, therapeutically promising parametric combinations is highly simplified. The parametric  continuum is discretized and the search is
conceived as the grid optimization.
For a specified discretization of the parametric space, eventual  inaccuracies may
result from the numerical instabilities 
or stiffness of the respective ODE integrators \cite{Butcher2008}.

As will be shown later, the optimality criteria are implicitly 
associated with the finite time interval, as the time horizon plays an important role in the decision-making.
Moreover, to guarantee that the optimization selects 
for the solutions with controllable sparsity \cite{Lethi2015}, 
the equality constraint or penalization term have been considered.
The main concepts of the approach may be summarized 
as follows:
\begin{enumerate}
\item The analysis of indirect influence of 
the cellular environments represented by the dynamics of the population numbers $B(t)$ and $C(t)$ (or the respective rates $R_{B}, R_{C}$)
on the tumoral populations quantified by $T(t)$, $J(t)$. 
The scheme shown in Fig.\ref{fig:ff1} outlines the 
topology of interactions and responses of $T$, $J$ to the exogenous parameters. 
Instead of analyzing intuitively obvious anti-proliferative/apoptotic 
effects controlled by the parameters $\alpha_C$, $\alpha_B$, $\beta_C$, $\beta_B$
we investigate sensitivity to the components of the tuple
\begin{equation}
\boldsymbol{\chi}_{\mathcal E} = \left[\,
h_{CT},\, g_{CC},\, g_{CB}, \,h_{BT}, \,g_{BC}, \, 
g_{BB} \,\,\right] \,,
\label{eq:listpar}
\end{equation}
which parametrize $(1+h_{CT} T) C^{g_{CC}}$, $B^{g_{CB}}$ and
$(1-h_{BT} T) C^{g_{BC}} B^{g_{BB}}$. (Here, the lower index $\mathcal{E}$ of $\boldsymbol{\chi}_{\mathcal E}$ emphasizes 
the environmental dependence).

\item The dynamical instability generating
tumor growth is induced by the super\-threshold 
choice of the parameter  $g_{TC}$ (see Table~\ref{tab:ParamX}). 
(The parameter appears in the term 
$\sim C^{g_{TC}} T^{g_{TT}}$  which constitutes $R_T$.)

\item 
The constancy of the selected parameters listed in the tuple
[Eq.~(\ref{eq:listpar})] 
is no longer considered, but the continuous in time dependencies 
will be introduced, where the original parameters are replaced by some 
exogenous time-varying functions
[see Eq.(\ref{eq:tildhCT}) in the text below]
{\small 
\begin{equation}
\tilde{\boldsymbol{\chi}}_{\mathcal E} (t,.) 
= \left[
{\widetilde h_{CT}}(t,.),\, 
{\widetilde g_{CC}}(t,.),\, 
{\widetilde g_{CB}}(t,.),\,
{\widetilde h_{BT}}(t,.),\, 
{\widetilde g_{BC}}(t,.),\, 
{\widetilde g_{BB}}(t,.)\,\, \right]
\label{eq:partim}
\end{equation}}
purposefully constructed to stay in the vicinity of their respective former static values.
The proposal of the specific time dependencies
implies the need for additional (discrete) parameters that not only
control the switchover 
between the base functions but undergo optimization as well.
Consequently, the constant parameters involved in $R_B, R_C$ 
are replaced by the corresponding time dependencies, formally
{\small 
\begin{eqnarray} 
R_C (\ldots ,\,  h_{CT}, g_{CC}, g_{CB}) 
&\longrightarrow & 
\widetilde{R}_C (\ldots ,\,  
{\widetilde h_{CT}}(t,.), \,\,{\widetilde g_{CC}}(t,.), {\widetilde g_{CB}}(t,.)) 
\,,\label{eq:replaceR}
\\
R_B (\ldots ,\,  h_{BT}, g_{BC},g_{BB}) 
&\longrightarrow &  \widetilde{R}_B (\ldots ,\, 
{\widetilde h_{BT}}(t,.), 
\,\,\, {\widetilde g_{BC}}(t,.), 
{\widetilde g_{BB}}(t,.)) \,.
\nonumber 
\end{eqnarray}}

\item The inverse formulation is suggested 
which assumes the construction of the objective function 
$f(.)$ which quantifies the degree of undesired behavior 
of $T(t)$ and $J(t)$. It follows that the aim of the optimization of $f(.)$ 
is the selection of Eq.(\ref{eq:partim})
which admits tumor suppression for given 
time interval. 

\end{enumerate}

\subsection{Discretization of the parametric space, exogenous symmetric pulses} \label{sub:QPsit}

The sensitivity analysis is based on the optimization of the exogenous stimulation 
from Eqs.(\ref{eq:partim}), (\ref{eq:replaceR})
which must be specified in more detail. 
In the present form it includes
combinations of the auxiliary discrete 
parameters forming the strings
\begin{eqnarray}
{\bf s} \equiv \left[ 
s_I, s_{h_{CT}}, s_{g_{CC}}, s_{g_{CB}}, s_{h_{BT}}, 
s_{g_{BC}}, s_{g_{BB}} \right]\,. 
\label{eq:svec}
\end{eqnarray}  
The grid optimization passes through the 
space of possibilities, where each possibility is encoded by the string ${\bf s}$.    
Continuity in time is guaranteed since 
two input pulses of prescribed shapes 
$\Psi_C(t,s_I)$, 
$\Psi_B(t)$ modify the original Eq.(\ref{eq:listpar})
in a multiplicative manner as it follows
\begin{eqnarray}
{\widetilde h_{CT}}(t, s_{h_{CT}},s_I,Q) 
&=& h_{CT}  \left(\, 
1+ Q s_{h_{CT}} 
\Psi_C(t, s_{I})\,\right)\,,
\label{eq:tildhCT}
\\
{\widetilde g_{CC}}(t,s_{g_{CC}},s_{I},Q) &=& 
g_{CC}  \left(\, 1+ Q s_{g_{CC}} 
\Psi_C(t, s_{I})\,\right)\,, 
\nonumber
\\
{\widetilde g_{CB}}(t,s_{g_{CB}},s_{I},Q) &=& g_{CB}  \left(\, 
1+ Q s_{g_{CB}} 
\Psi_C(t, s_{I})\,\right)\,, 
\nonumber
\\
{\widetilde h_{BT}}(t,s_{h_{BT}},Q)  &=&
h_{BT}  \left(\, 
1 +  Q s_{h_{BT}} \Psi_B(t) \,\right)\,,
\nonumber 
\\
{\widetilde g_{BC}}(t,s_{g_{BC}},Q)  
&=& g_{BC}  \left(\, 1 +  Q s_{g_{BC}} \Psi_B(t)\,\right)\,, 
\nonumber
\\
{\widetilde g_{BB}}(t,s_{g_{BB}},Q)  
&=&  g_{BB}  \left(\, 1 +  Q s_{g_{BB}} 
\Psi_B(t) \,\right)\,.
\nonumber
\end{eqnarray} 
The parameter $Q\geq 0$ may be interpreted 
as the strength of therapeutic action. 
In the modeling below, we focus on the possible shapes of multi-dimensional pulse
and their responses which can be used to model reversible variations
of individual parameters. For that purpose, the discrete argument
$s_I$ of $\Psi_C (t,s_I)$ is introduced [see Eq.(\ref{eq:PsiBC})
and illustrating Fig.\ref{fig:ff2} below] to control timing and width 
of the pulse. Regarding the focus of our study, continuous (constant) administration is 
excluded because of the requirement of dominating autonomous regimes in the late-time systemic 
response which is equivalent to the condition of the 
presence of a rest period that allows for the recovery from 
toxicity \cite{Foo2009}.

The space of pulse configurations is determined by ${\bf s}$ 
including the four-state component $s_I \in \Omega^{(4)}\myeq \{0,1,2,3\}$,
and six three-state components $s_{index} \in $ $ 
\Omega^{(3)} \myeq \{-1,0,1\}$, where
$index \in \{$  
$ h_{CT}$, 
$ g_{CC}$, $g_{CB}$, $ h_{BT}$, 
$ g_{BC}$, $g_{BB}\}$. 
Next we will study optimality 
of the strings 
\begin{equation}
{\bf s} \in \Omega \myeq 
\Omega^{(4)} \times  
\underbrace{
\Omega^{(3)} 
\times \Omega^{(3)} \times \ldots \times  \Omega^{(3)}}_{6\times}\,,
\label{eq:Om6x}
\end{equation}
where the cardinality $card(\Omega)=4\times 3^6=2916 $ possible states.  

The  functions $\Psi_C(t,s_I)$, $\Psi_B(t)$  
used to carry out the sensitivity testing
are modeled by means of the elementary 
symmetric pulses
\begin{eqnarray}
\label{timing}
\Psi_B(t) &=& 
\psi(t,t_0,\xi_0,\sigma_0)\,, 
\label{eq:PsiBC}
\\
\Psi_C(t,s_I) 
&=& \delta_{s_I,0}\, 
\psi(t, t_0,\xi_0,\sigma_0)+
\delta_{s_I,1} \,
\psi(t, t_0,\xi_1,\sigma_0)
\nonumber
\\
&+&\delta_{s_I,2}\, 
\psi(t, t_1,\xi_0,\sigma_0)+
\delta_{s_I,3}\, 
\psi(t, t_1,\xi_1,\sigma_0)\,,
\nonumber
\end{eqnarray}
where $\delta_{s_I,s}$ denotes the Kronecker symbol, 
which equals to $1$ or $0$ depending on the match or mismatch status of 
$s_I$ and one of the referential values from $ \Omega^{(4)}$.
Note that the pulse timing and shaping encoded 
by $s_I \in \Omega^{(4)}$ undergo optimization as well.
Incorporation of the sequential pulses controlled 
by $s_I$ is inspired by \cite{Nichol2015}.
  
We remind that the role of the parameter $s_I$ is crucial,
since it determines the timing and shape of the pulses; 
$\Psi_B(.)$ is not affected by this parameter and serves as
referential.
For all possible pairs of arguments $( t_0,\xi_0)$;  \,
$( t_0,\xi_1)$; \,
$( t_1,\xi_0)$; \, $(t_1,\xi_1)$
including the time-translation $t_{tr}\in \{t_0, t_1\}$ 
and time rescaling factors $\xi_{sc}\in \{\xi_0,\xi_1\}$ we define
four child wavelets [see Eq.(\ref{eq:PsiBC})] 
\begin{equation}
\psi(t, t_{tr}, \xi_{sc},\sigma_0)= 
\frac{1}{\sqrt{ \xi_{sc}}}
\phi\left(\frac{t-t_{tr}}{\xi_{sc}},\sigma_0\right)
\label{eq:psi}
\end{equation}
which exploit the continuous-time univariate Gaussian symmetric model
\begin{eqnarray}
\phi(t) =  \phi_S(t',\sigma_0)=
\exp\left[
-\frac{1}{2} 
\left(\frac{t'}{\sigma_0}\right)^2 \right]\,, 
\label{eq:sigmt}
\end{eqnarray}
where $\sigma_0$ 
is the pulse width [$\sigma_0<(t_1-t_0)$].

More realistic (asymmetric) variant of the function $\phi$ is introduced 
and studied in sec. \ref{sec:APulse}. 
To avoid the direct impact of the pulses on the boundaries, 
the times $t_0$, $t_1$ were chosen to guarantee the localization of the pulse peaks far enough from
the boundary values $t_S$ and $t_E$, where $t_E>t_1>t_0>t_S$, i.e. we assume 
$\psi\sim 0$ for $t \in \{ t_S, t_E\}$. The time span $t_E-t_S- 2\sigma_0$ 
corresponds, at the conceptual level, to what 
is called the "drug holidays" \cite{Cappucio2007,Castiglione2006,Foo2009}.
Unlike the superposition of Dirac pulses, which define the control function 
in the above-cited reference, 
we have the finite pulse width $\sigma_0$. 
Obviously, the symmetry of input signal does not imply of the population responses.
In particular, the unstable tumor growth causes irreversible changes which, in principal,
can not be compensated solely by the action of reversible exogenous factors selected here.
To summarize this section, the auxiliary functions 
$\Psi_B(.)$ and $\Psi_C(.)$
were introduced to construct the reversible exogenous pulses 
of dimension six [see Eqs.(\ref{eq:tildhCT}) and (\ref{eq:Om6x})]. 

\subsection{Objective function, constrained optimization of tumor response}

In this section we focus on the optimization of tumor responses
to the parametric environmental factors with the aim to identify those factors 
which lead to required changes of the tumoral 
subpopulations during some interval of observation.  
In conformity with the optimization theory, the objectives
of indirect parametric and environmental manipulations with tumoral populations  
are quantified by the objective function.
The proposed formulation consists of two main components. 
At first, the optimization component enables us to discriminate between many
alternative therapeutic strategies (and to select, regarding 
the model, the optimum solution).
Secondly, the component of sensitivity analysis
enables more subtle understanding of the relationships 
between exogenous inputs and output variables 
in a system and, in addition, it
enables to study degree of sparsity 
and robustness \cite{Castiglione2006} of the 
proposed solutions.
To pursue the above aims in our specific case 
of the model of unstable tumor growth, several key measures 
of tumor response are incorporated
into the objective function.

\subsubsection{The measures of tumor responses, objective functions}\label{sec:tumrespons}

Below we introduce the key measures that reflect tumor responses 
in our specific case of unstable tumor growth. The time 
interval for the observation of the systemic dynamics
is $\langle\, t_S, t_E\,\rangle$ ($t_S$ stands for the time of start, $t_E$ for the end).
Further, we assume that population variations within the 
interval $\langle\, t_S, t_E\,\rangle$
do not contribute directly to the objective functions 
values that are calculated purely from the 
population characteristics at the interval endpoints. 

Manipulation with several tumor populations requires 
multi-objective formulation. As the optimality should be evaluated
not only for $T$ but for $J$ as well, two different 
objective functions are introduced  
 \begin{equation} 
f_{T}(t_S,t_E,{\bf s}) \myeq  \frac{T(t_E,{\bf s})}{T(t_S)}\,,
\qquad 
f_{J}(t_S, t_E, {\bf s}) \myeq \frac{J(t_E,{\bf s})}{J(t_S)}\,, 
\label{eq:fTJ}
\end{equation}
where the notation emphasizes the conditioning by discrete ${\bf s}$.
We formulate the objective function in the terms of relative (instead of
absolute) tumor abundances to be able to stabilize non-free tumor equilibria
by environmental moves (e. g. the pulses of therapy). 
Such formulation of the objective function is in line 
with the concept of adaptive therapy \cite{Gatenby2009b}.
The alternative formulation exists \cite{Qiao2015} based on the OBs and OCs abundances  
that aims to restore their balance disrupted by MM.

Productive and broadly accepted paradigm in the area of the multi-objective optimization is the 
Wald's "minimax" optimality \cite{Wald1950,Parmigiani2009,dAlbis2012}.
The scalarization procedure was designed to solve multi-objective 
decision-making problems where the decisions are made on the basis of the worst possible choice. 
Owing to this, both measures from Eq.(\ref{eq:fTJ}) 
can be incorporated into the single objective function form 
\begin{equation}   
f(t_S,t_E,{\bf s})  \myeq
\max \{\, f_{T} (t_S,t_E,{\bf s}),\, f_{J} (t_S,t_E,{\bf s})\,\}\,,
\label{eq:maxf}
\end{equation}
where the optimal string ${\bf s}={\bf s}_{min}$ is defined in the standard way
\begin{equation}
{\bf s}_{min} \myeq 
\arg\min_{ {\bf s} \in \Omega} f(t_S,t_E,{\bf s}) \,.  
\label{eq:smin}
\end{equation}

At the moment, we do not incorporate metastatic 
potential and drug resistance consequent to the JCs 
\cite{Bouchnita2016,Tang2016} into the model.
In the future, an eventual metastatic population derived 
from the JCs could be incorporated as 
the third component of the objective function.

In order to control the number of non-zero ${\bf s}$ components ($+1$ 
corresponds to stimulation, while $-1$ stands for inhibition), 
the optimization task from Eq.(\ref{eq:smin}) is supplemented 
with the constraint
\begin{equation}
\delta_{0, s_{h_{CT}}} + \delta_{0, s_{g_{CC}}}+
\delta_{0, s_{g_{CB}}} + 
\delta_{0, s_{h_{BT}}} +
\delta_{0, s_{g_{BC}}} + 
\delta_{0, s_{g_{BB}}}= n_0({\bf s})\,,
\label{eq:n0}
\end{equation}
where $n_0({\bf s})=0, 1, \ldots, 6$ is the number of ${\bf s}$ components that are equal to $0$; 
$\delta_{s_x, s_y}$ is the Kronecker symbol. The variable $s_I\in \Omega^{(4)}$ absents in the 
l.h.s. of Eq.(\ref{eq:n0}) as it relates only to the timing and width 
of the pulses (Eq.(\ref{timing})).

As up-to-date anticancer therapies are accompanied by the
variety of negative or uncertain 
side effects, the constraints may represent the 
elementary  quantification of their impacts \cite{Foo2009}. 
When $6-n_0$ increases, the tendency 
of the system towards more complex
response becomes unavoidable. 
Therefore, because of direct link between the number of 
constraints and parametric sparsity, 
the term $1-(n_0/6)$ may be interpreted as the index of  
{\em parametric} {\em redundancy} \cite{Deun2011}. 
More compact, constraint-free reformulation where adverse 
effects are quantified by $1-(n_0/6)$ regularization term 
is presented in sec.~\ref{se:lamPen}.

\subsection{Evaluation of the model outputs}

The optimum value of the objective function  
\begin{equation}
f_{min} \equiv f_{min}(t_S,t_E) \myeq\min_{{\bf s}\in \Omega} f(t_S,t_E,{\bf s})\
= f(t_S,t_E,{\bf s}_{min})
\end{equation}
enables straightforward quantification of the quality of the optimization outputs. 
The disparities in the optimization of $f_T$ and $f_J$ are represented 
by the measure
\begin{equation}
Df_{min} \myeq  \Big|\, 
\min_{{\bf s} \in \Omega}  f_{T}   -  
\min_{{\bf s} \in \Omega}  f_{J}\, \Big|\,. 
\label{eq:ADf}
\end{equation}
In addition, to understand how the choice of the fitness 
variant affects ${\bf s}_{min}$,  we introduced 
the mean Hamming distance 
\begin{equation}
d_{H,smin} \myeq \frac{1}{7}\, 
\sum_{\,\,\,\,\,\,\,\{\forall \, {\bf s} \,\, \mbox{\tiny components}\}} 
\mbox{\Large ${\mathds{1}}$}
\left(
\arg \min_{s\in \Omega} f_T 
\neq 
\arg \min_{s\in \Omega} f_J\,
\right)\,,    
\label{eq:dH} 
\end{equation}
where ${\mathds{1}}(.)$ is the indicator 
function of the logical-type argument, 
which returns one when the argument is True and zero 
otherwise. Having defined a per-component distance measure,
we consider the normalization factor $1/7$. 

\section{Numerical implementation, symmetric pulses}\label{sec:num}

\subsection{Parameter settings}

The parameters used in all simulations 
are consistent with those used by Koenders and Saso \cite{Koenders2016}.
Their values are listed in Table~\ref{tab:ParamX}.
Additional in our approach are the parameters which 
define the exogenous dynamics of the functions  
$\psi(t,t_{tr},\xi_{sc},\sigma_0)$, $\phi_S(t,\sigma_0)$ 
[see Eq.(\ref{eq:psi}) and Eq.(\ref{eq:sigmt})] which provide 
$\Psi_B(t)$, $\Psi_C(t,s_I)$ [see Eq.(\ref{eq:PsiBC})]
for given $s_I$; their values are
\begin{equation}
t_0 =140 \,day  >  t_S= 100\, day, 
\,\,\,\,
t_0 < t_1= 160 \,day < t_E= 200\, day
\label{eq:paramday}
\end{equation}
and the scaling parameters needed for the evaluation of the functions 
\begin{equation}
\xi_0 = 1, \,\, \xi_1=0.5  \,\, 
(\mbox{\small scaling\, factors})\,,
\quad  
\sigma_0=7\, day\,
(\mbox{\small pulse\,width})\,.
\label{eq:sigmaday}
\end{equation}
The pulse width $2\sigma_0=14 day$ has been chosen to be roughly 
consistent with the half-life of MM cells that is $\sim 10-20$ days. 
The choice $t_E-t_S=100 day$ was primarily motivated 
by elsewhere referred  average time for the restoration 
of the population size towards equilibrium 
\cite{Komarova2003,Koenders2016}. The importance 
of this time scale is supported by the 
work \cite{Ayati2010}.

The overall dynamics and system responses were 
obtained by means of the fourth-order 
Runge Kutta (RK4) method with 
the integration 
step $\delta t=5 \times 10^{-4} day \sim 8.64 \,sec$. 

\begin{table}[bth]
\begin{tabular}{||l|l||l|l||l|l||}
\hline
{\small param. val.}   & {\small class}  & 
{\small param. val.}    & {\small class} & 
{\small param. val.}    & {\small class}  \\
\hline
\hline
$\alpha_C= 3.0\, day^{-1}   $             &    $R_C$  & 
$\alpha_B= 4.0\, day^{-1}   $             &    $R_B$  &
$\alpha_T= 0.3\, day^{-1}   $             &    $R_T$  \\ 
$\alpha_J= 0.001 \,day^{-1} $           &    $R_J$  &
$\beta_C= 0.2 day^{-1}      $           &    $R_C$  &
$\beta_B=  0.02 \,day^{-1}  $           &    $R_B$  \\ 
$\beta_T = 0.1\, day^{-1}   $           &    $R_T$   &
$\beta_J = 0.3\, day^{-1}   $           &    $R_J$   &
$\kappa_J = 0.3 \,day^{-1}  $           &    $R_T$  \\
$\underline{h_{BT}= 0.035}  $    &  $R_B$       &    
$\underline{h_{CT}= h_{BT}} $    &  $R_C$      &   
$\underline{g_{CC}= 0.5}  $      &  $R_C$      \\    
$\underline{g_{CB}= -0.5} $      &  $R_C$     &   
$ \underline{g_{BC}= 1.0} $      &  $R_B$      & 
$ \underline{g_{BB}= 0.0} $       &  $R_B$      \\
$ \underline{g_{TT}= 0.5} $        &  $R_T$      &
-------------- & ------  &
$ g_{TC,init}=0.0$                &  $R_T$         \\
$ g_{TC,stab}=0.3$              &  $R_T$           &
$ g_{TC,thr}=0.3648$           &  $R_T$            &
$ g_{TC,unstab}=0.37$         &  $R_T$             \\  
\hline 
\end{tabular}
\caption{The constant 
model parameters used in all simulations. 
The parameter values ({\small param. val.}) are adopted 
from the work \cite{Koenders2016}.
They are divided into four classes - four rates,
$R_C, R_B, R_T, R_J$ according to the occurrence
on the right hand side of Eq.(\ref{eq:RCRBRTRJ}).
The underlined parameters 
$h_{BT},\ldots, g_{TT}$ are later replaced by their 
corresponding 
time-varying analogs [see Eq.(\ref{eq:tildhCT})]. Four variants of the 
parameter $g_{TC}$ (bottom rows of the table)
were used to simulate the stability/instability 
of steady states in the tumor growth context.}
\label{tab:ParamX}
\end{table}

\subsubsection{Static equilibrium - stability}\label{sec:eqstab}

Before going into the details of the parameter settings in dynamic approach,
we briefly discuss some of the 
static results. The approximate population equilibrium 
$B=B_{eq}^{(0)}=230.87,$  
$C=C_{eq}^{(0)}=1.685,$ 
$T=T_{eq}^{(0)}=9.0$ was calculated 
using Eqs.~(\ref{eq:scalform1}), (\ref{eq:scalform2}), 
adopted from \cite{Koenders2016}. 
We note that the "zero-order" 
equilibrium did not take into account population
of the JCs $(J=0)$. In addition, the 
approximation assumed $g_{TC}=g_{TC,init}=0.0$. 
By using asymptotes 
of direct integration of Eqs.(\ref{eq:CBTJ}) and (\ref{eq:RCRBRTRJ}) 
for appropriately chosen nonzero constant 
value $g_{TC}=g_{TC,stab}=0.3$, 
the previous equilibrium estimates changed
and the asymptotically 
stable equilibrium $B^{(1)}_{eq}=211.42,$  
$C^{(1)}_{eq}=2.052,$  
$T^{(1)}_{eq}=13.854$, $J^{(1)}_{eq}=0.0947$
was obtained. 

\subsection{Initiation, imbalance, pathogenesis and MM progression}\label{sub:Init}

Our modeling of initiation and promotion of MM pathogenesis takes into account 
early steps of OCs involvement in that process. According to \cite{Yaccoby2004}, 
the direct interactions between MM cells and OCs may increase MM proliferation which 
is mediated by $\sim C^{g_{TC}} T^{g_{TT}}$ term.  Therefore, we assume 
that the large pathogenic irreversible changes can be modeled with increased 
$g_{TC}$ which causes unstable growth of the populations of MM cells. 
The parametric shift in $g_{TC}$ exceeding $g_{TC,stab}$ results in the permanent 
instability that may be attributed to the abnormal cell signaling, 
modulation of microenvironment, or dysregulation that 
irreversibly degrades bone. From the point of view of the evolutionary game theory 
\cite{Dingli2009}, the shift can be attributed to the perturbation of OC-MM 
coexistence line. A more microscopic interpretation can be found in
\cite{Edwards2008,Qiao2015}, where the MM growth is stimulated by OCs 
secretion of TNF$_\alpha$ cytokine.

We have numerically found that the parameter $g_{TC}$ reveals the threshold value
$g_{TC,thr} \simeq 0.3648$ which separates the stable equilibrium from 
the unstable regime. Consequently, the supercritical value 
$g_{TC,unstab}=0.37$ $ > g_{TC,thr}$ was used in the tumor 
growth simulations. The unstable system is integrated over the initial 
$t_S=100 day$ with constant parameters, i.~e. for $Q=0$. 
This period is used to simulate the uncontrolled growth. 
The resulting values $T(t_S)=16.784$ and $J(t_S)=0.132$ 
have been stored for later comparative purposes.  
After this period, the tumor suppression is purposefully 
initialized. Subsequently, the equations were 
integrated for the time interval 
$\langle t_S, t_E \rangle$, where $t_E=200\,day$.
The pairs $( T(t_S),\, T(t_E))$, 
$(J(t_S), J(t_E))$ were used 
to calculate the objective function defined by 
Eq.(\ref{eq:fTJ}). In the case $Q=0$ (without pulses), 
when there is no dependence on ${\bf s}$ (total degeneracy)
and, consequently, no optimization is possible,
the calculation provides $f=f_J = 1.1568 > f_T=1.0681$. 
This result suggests that, at least in the low $Q$ region, 
more rapidly expanding population of the JCs
could be suppressed with higher priority [in the light of the criterion 
specified in Eq.(\ref{eq:maxf}). 

\subsection{Results, symmetric pulses}\label{sub:sympRes}

\begin{figure}[t]
\includegraphics[width=0.92\linewidth]{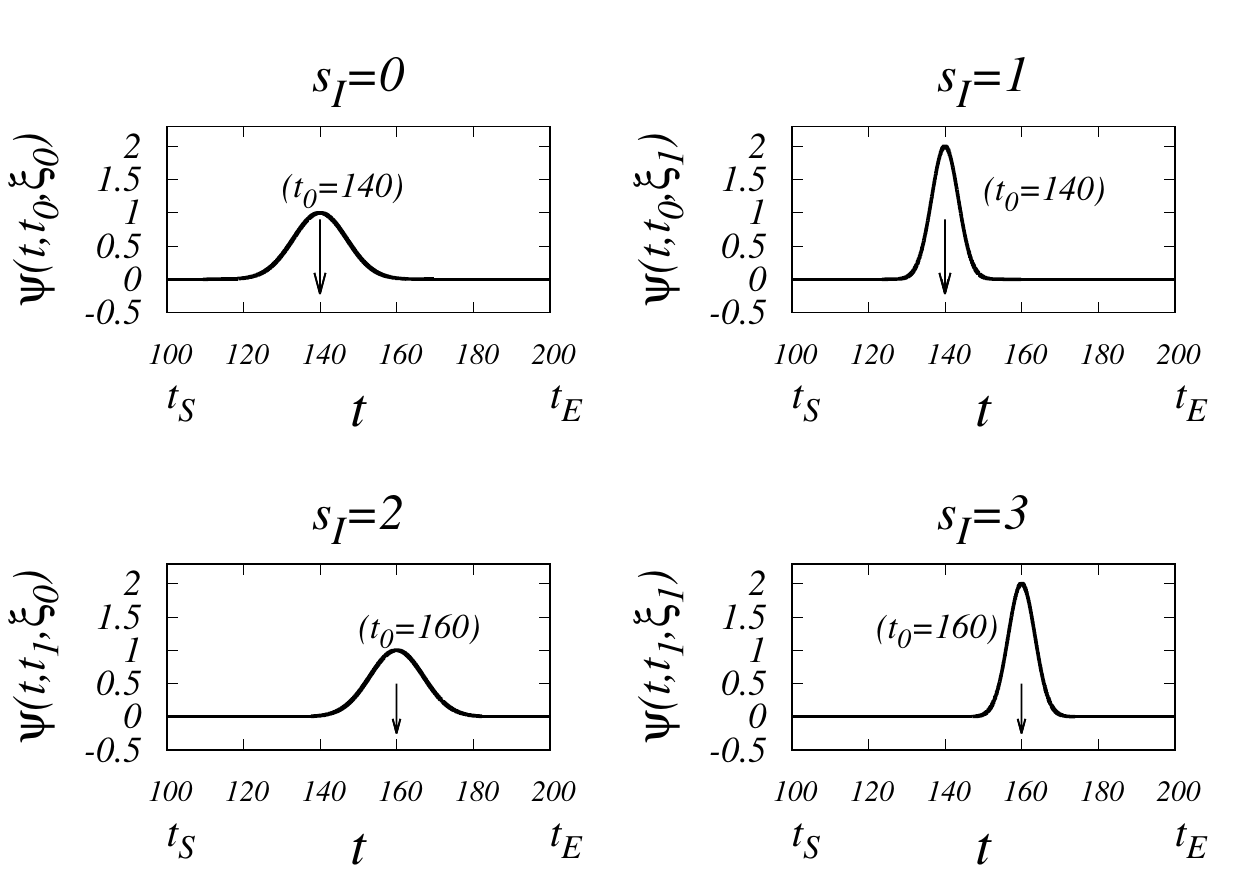}
\caption{The dynamic elements 
of the sensitivity analysis generated 
by the translation and scaling of 
the arguments of the Gaussian pulses from Eq.(\ref{eq:sigmt}). 
The calculation is based on the parameters $t_0, \xi_0, t_1$, 
$\xi_1, \sigma_0$ defined by Eqs.(\ref{eq:paramday}) 
and (\ref{eq:sigmaday}). The positions 
of $t=t_0$ are indicated by the vertical arrows. 
The examples of four child wavelets Eq.(\ref{eq:psi})
used to construct $\Psi_B(t)$ (corresponding to $s_I=0$) and 
$\Psi_C(t,s_I)$ (for $s_I\in \Omega^{(4)}$).}
\label{fig:ff2}
\end{figure}

\begin{figure}[bth]
\mbox{
\begin{tabular}{ll}
\includegraphics[width=0.7\linewidth]{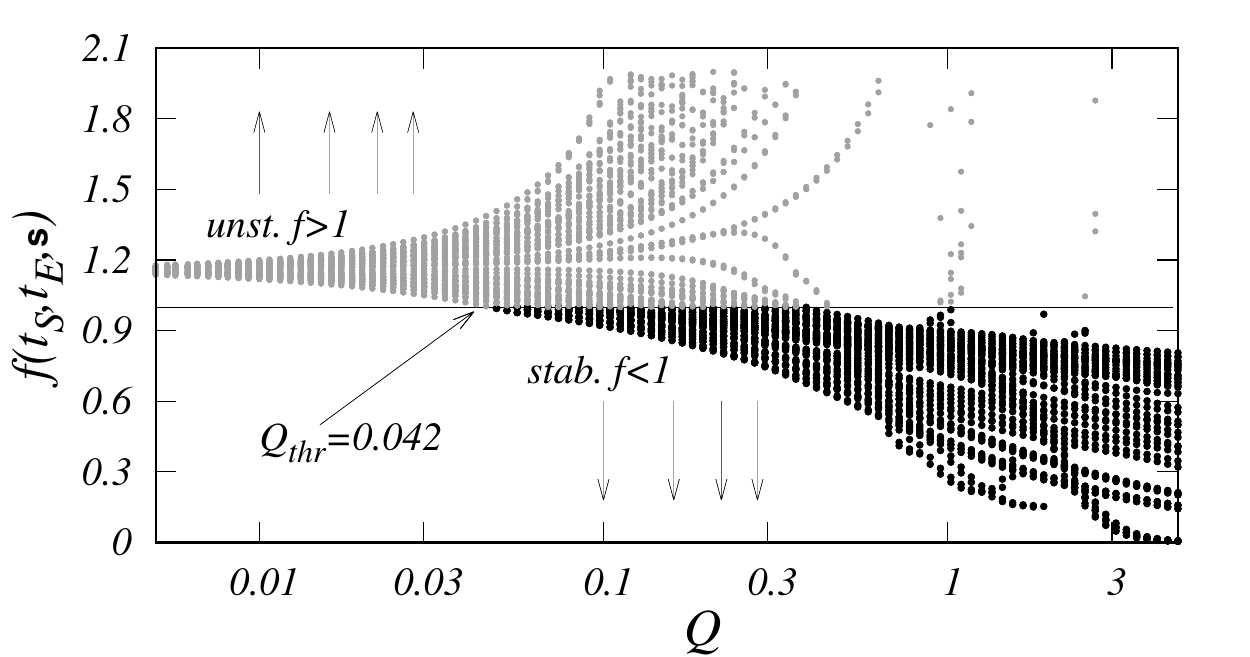}
&
\hspace*{-10mm}
\includegraphics[width=0.3\linewidth]{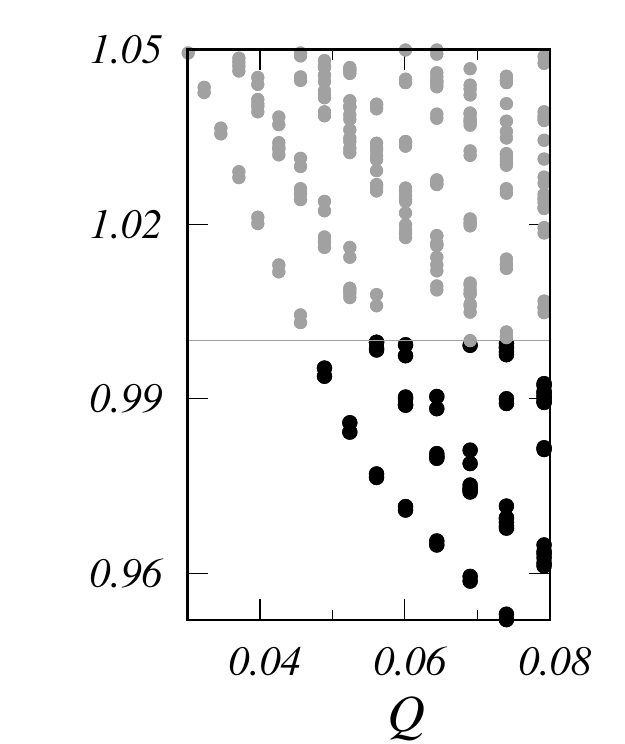}
\\
\vspace*{-5mm}
\\
\mbox{(a)} \mbox{$n_0=0$}  & 
\hspace*{-10mm}
\mbox{(b)} \mbox{$n_0=0$ detail}
\\
\vspace*{-1mm} &
\\
\includegraphics[width=0.7\linewidth]{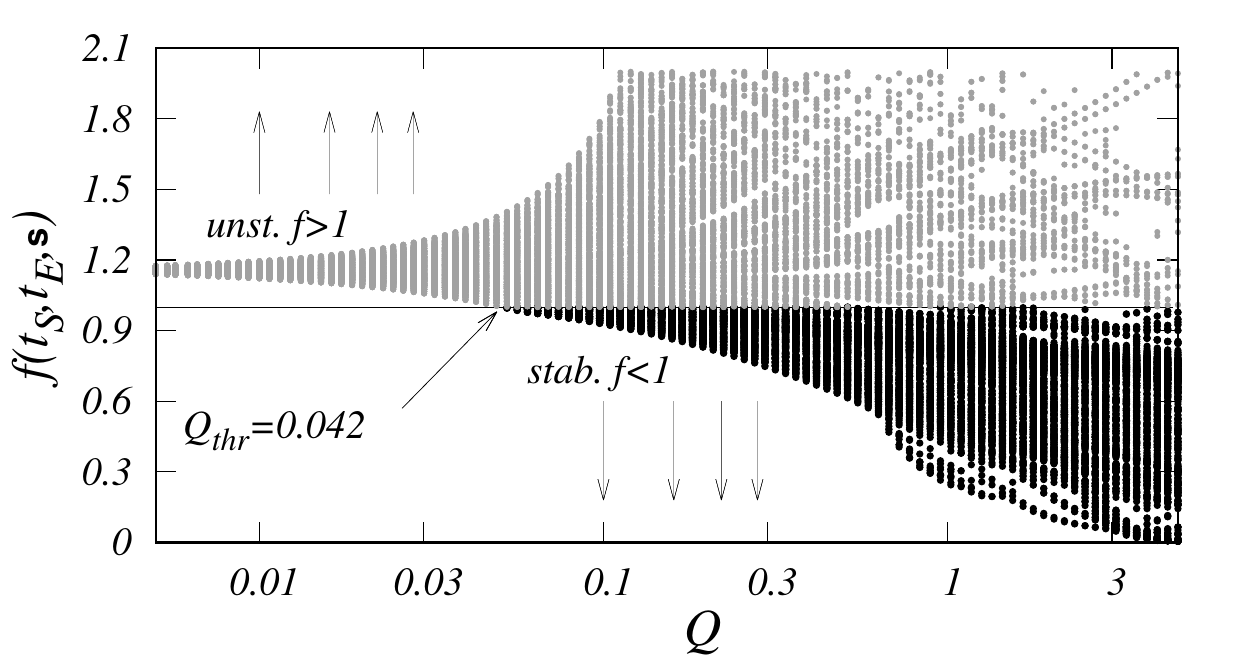}
&
\hspace*{-10mm}
\includegraphics[width=0.3\linewidth]{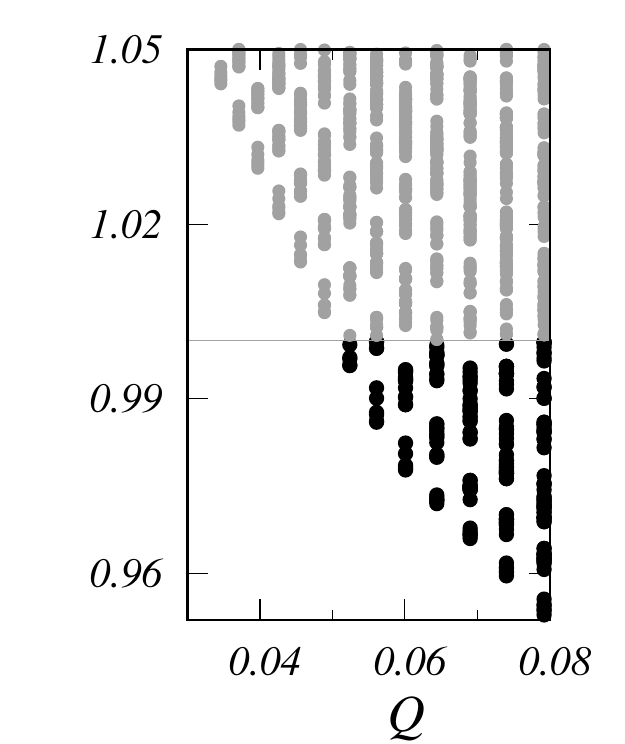}
\\
\vspace*{-5mm} &
\\
\mbox{(c)} \mbox{$n_0=2$} &  
\hspace*{-10mm}
\mbox{(d)} \mbox{$n_0=2$ detail}
\\
\end{tabular}}
\caption{
The motivation for the optimization. 
The figure panels calculated for different constraints,
$n_0=0$ (panels (a), (b)) and $n_0=2$ (panels (c), (d)) 
The values of the objective function $f(t_S,t_E,{\bf s})$ 
are calculated for all  strings ${\bf s}\in \Omega$ and 
plotted as the function of $Q$. 
The results reveal 
the existence  of parametric (and constraint dependent) threshold 
$Q_{thr}(n_0)$ above which the therapeutic effect $f(.)<1$ 
exists when the optimization is applied. 
The details of the threshold 
neighborhoods are shown in the panels (b), (d). 
In all cases the stability line $f=1$ separates unstable (unst.) from 
stable (stab.) regions. The highly unstable 
region $f>2$ was removed from the plot.}
\label{fig:ff4}
\end{figure}

Before presenting the simulation results, 
we visualize examples of the elementary Gaussian exogenous pulses
[see Eq.(\ref{eq:PsiBC}), Fig.\ref{fig:ff2}] for different $s_I$. 
Next, instead of immediate quest for the optimality,
we find instructive to inspect the entire search space. 
Our findings (see Fig.~\ref{fig:ff4}) demonstrate that  not all the 
strings ${\bf s}\in\Omega$ are tumor suppressive,
i.~e. providing $f(.)<1$. It seems that the outputs quantified by $f(.)$ 
split into qualitatively different branches.
One may identify the options 
${\bf s}\in \Omega$ corresponding to 
the quasi-equilibrium (line $f=1$), 
unstable tumor growth ($f>1$), 
or temporary tumor suppression ($f<1$), respectively. 
It can be seen that for too small $Q$ the desired effects ($f<1$) 
can not be achieved, and growth instability may be suppressed only for
$Q>Q_{thr} =0.049$. It demonstrates that the parametrization covers 
a broad range of scenarios, so that optimization can affect the therapeutic efficiency.

\begin{figure}[bth]
\begin{tabular}{ll}
\includegraphics[width=0.49\linewidth]{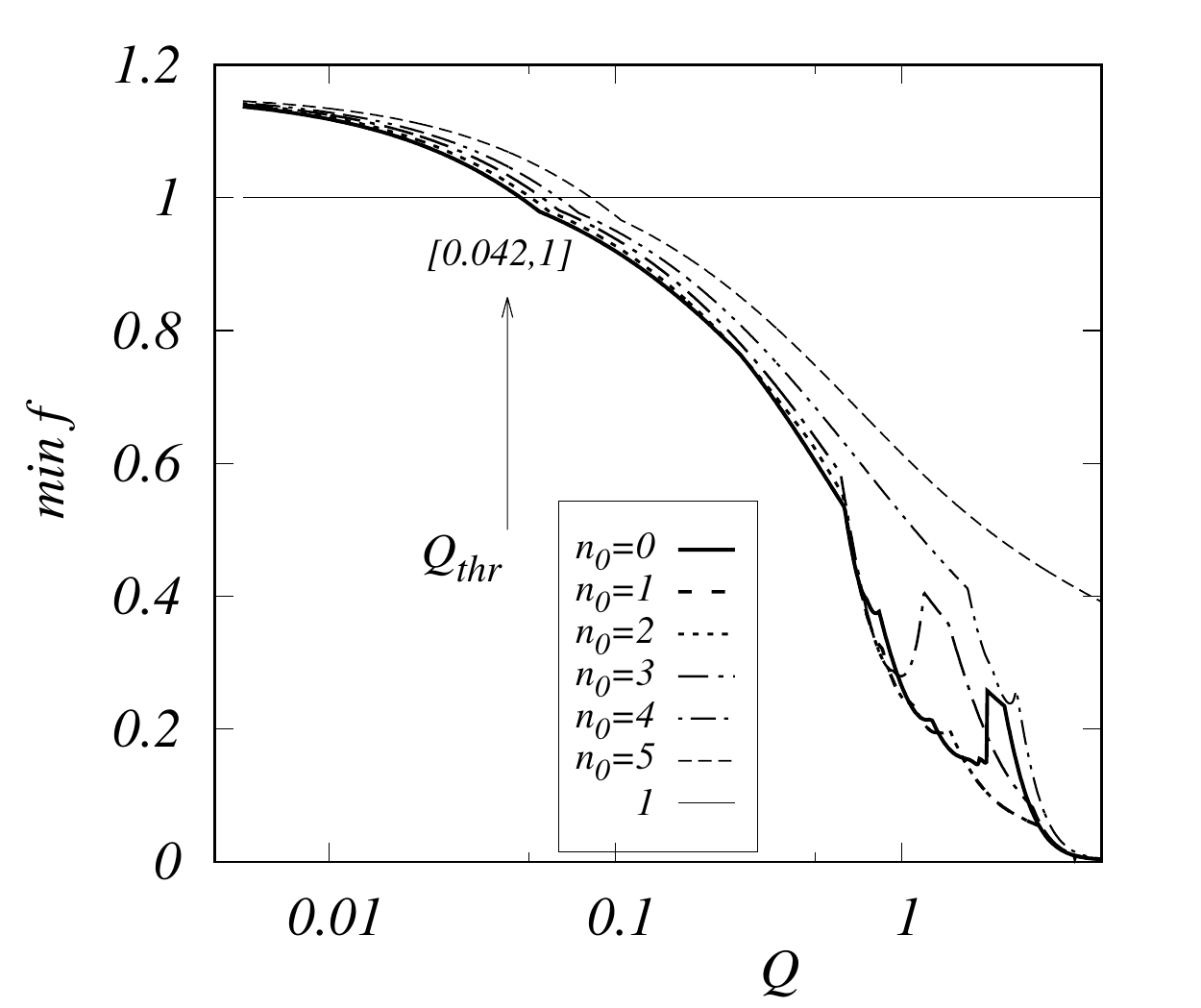} 
&   
\includegraphics[width=0.49\linewidth]{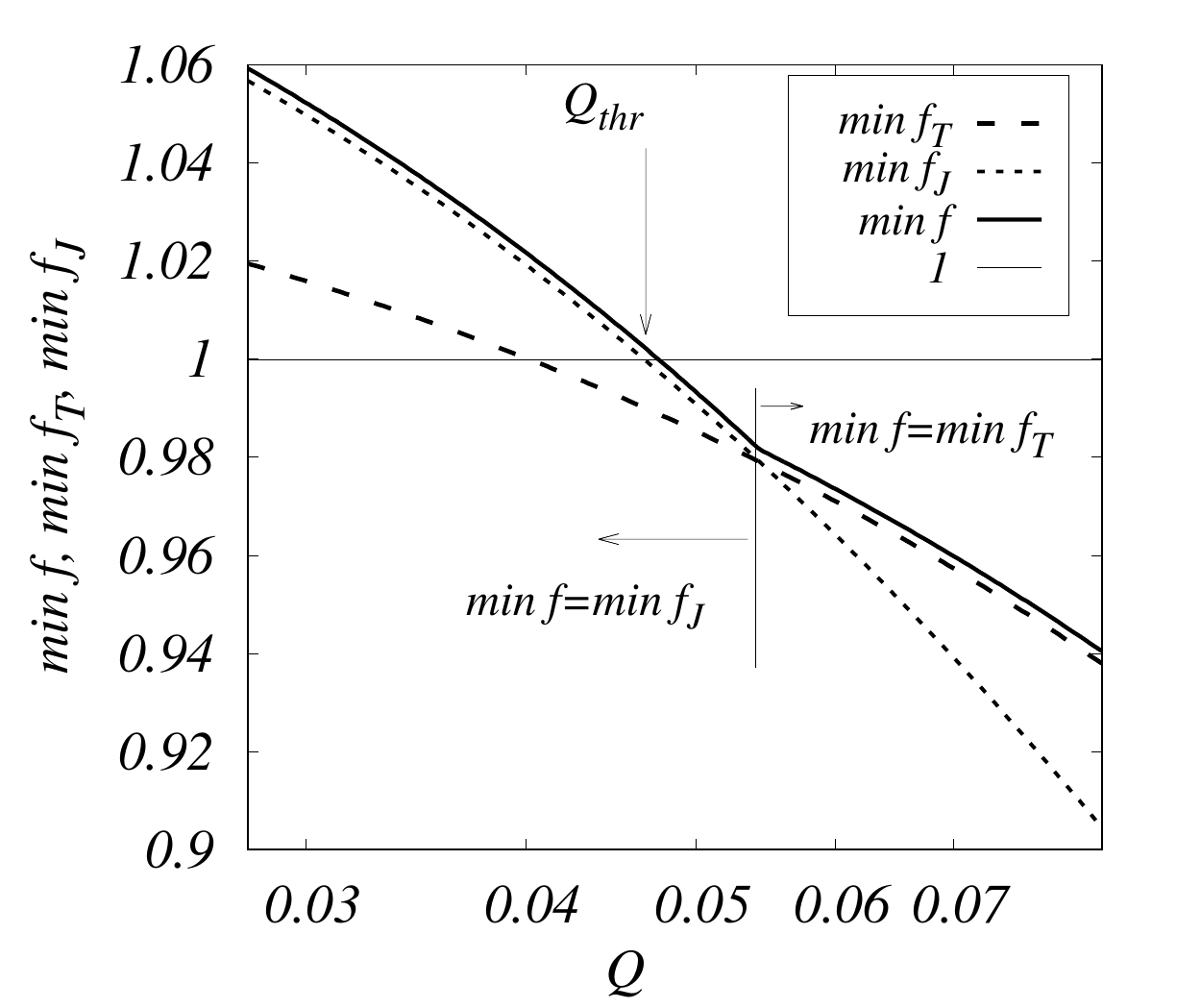}
\\
\vspace*{-7mm}
&
\vspace*{-4mm}
\\
\mbox{\scriptsize (a)} 
& 
\mbox{{\scriptsize (b)} {\scriptsize detail, $n_0=0$}} 
\end{tabular}
\caption{The plot of the optimized $f({\bf s})$. 
Panel (a) shows the $\min f$ dependencies 
on $Q$ calculated for six constraints 
$n_0=0, 1, \ldots, 5$. It means 
that strictly one discrete component among 
$s_{h_{CT}}$, $s_{g_{CC}}$, 
$s_{g_{CB}}$, $s_{h_{BT}}$, $s_{g_{BC}}$, 
$s_{g_{BB}}$ is requested 
to be nonzero in the case of $n_0=5$ constraint.  The result 
that higher $n_0$ leads to lower efficiency 
of the virtual therapies (especially for high $Q$) 
is in agreement with intuitive reasoning. 
However, the peaked form of the $f_{min}(Q)$ at high $Q$
lacks the intuitive understanding. 
It can be understood as a 
kind of the combined effect of the optimization, high 
amplitude stimuli and nonlinearities of ODEs system. 
Panel (b) sheds light on the problem of threshold 
$Q_{thr}$ where $f(Q_{thr})=1$.
The careful analysis shows that it is close 
but not identical to the point
$\min f_T = \min f_J$ where the objects 
of the optimization
 [larger function from the pair 
$f_T$, $f_J$, see Eq.(\ref{eq:maxf})] exchange.}
\label{fig:ff5}
\end{figure}

\begin{figure}[bth]
\mbox{
\begin{tabular}{l}
\includegraphics[width=0.99\linewidth]{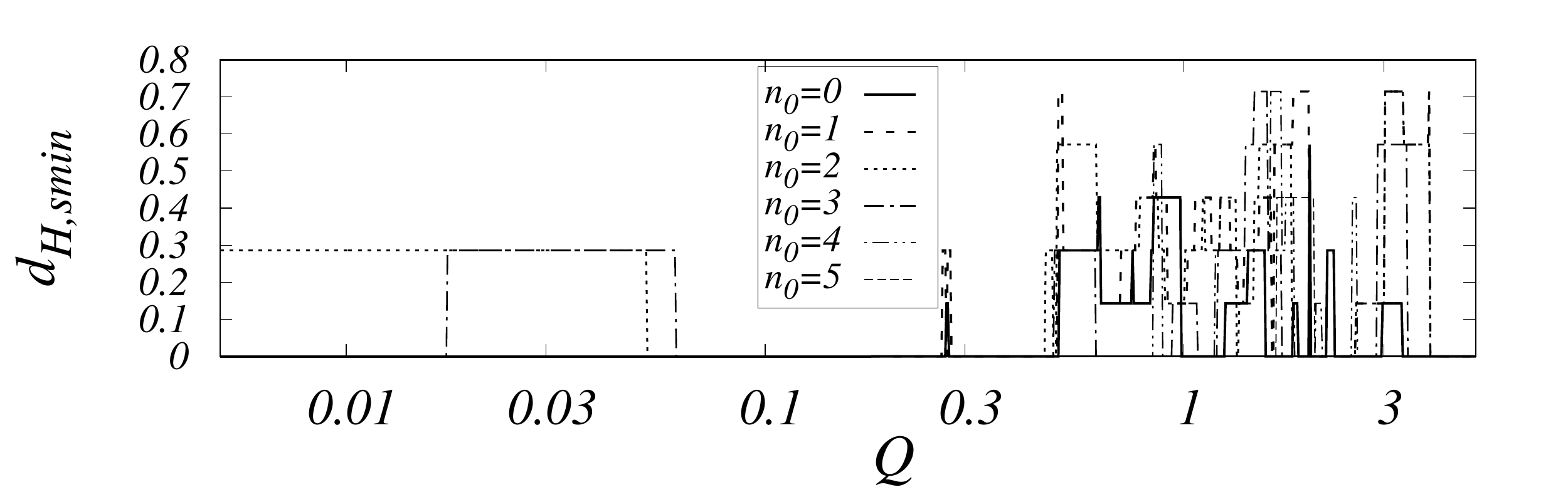}
\\
\vspace*{-12mm}
\\
\mbox{\small (a)} 
\\
\vspace*{-6mm}
\\ 
\includegraphics[width=0.99\linewidth]{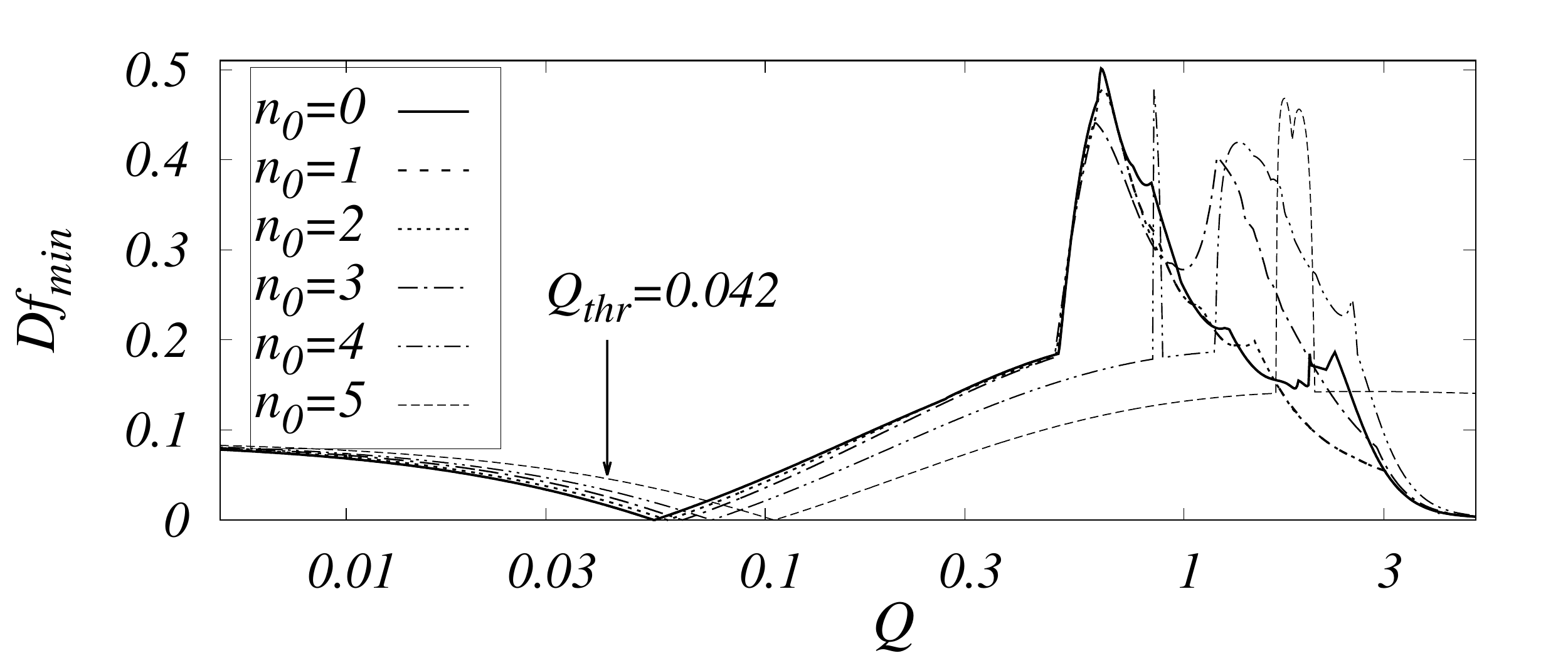}
\\
\vspace*{-12mm}
\\
\mbox{\small (b)} 
\end{tabular}}
\caption{
The panels show differences between 
the applied optimization criteria expressed by 
the functions $f_T$ and $f_J$ calculated  
for different $n_0$. The 
differences in the minima of  
$f_T$ and $f_J$ are quantified by means of the 
Hamming distance $d_{H,smin}$. 
The objective function values 
are reflected by the diversity 
measure $Df_{min}$ [see panel (b)] 
defined by Eq.(\ref{eq:ADf}). 
The subtle difference between the threshold position 
$Q_{thr}$ (for $n_0=0$) and 
the minimum of $Df_{min}$ obtained for 
$n_0=0$ has been already explained 
in the Fig.\ref{fig:ff4}.  
The highest diversity occurs at the large $Q$.}
 \label{fig:ff6}
\end{figure}

After illustrating all the possible scenarios contained in $\Omega$,
we turn to the analysis of the optimal selections (see Fig.\ref{fig:ff5}).
It seems that the small and large $Q$ regions differ in qualitative as well as
quantitative components. The complex form of $f_{min}(Q)$
consisting of several sharp folds arises at large $Q$. 
This is because of combined effect of discretization, 
selection, and nonlinearities, as well as the strength of exogenous factors.
With increasing $n_0$, the sharp irregular folds in the $f_{min}(Q)$ 
dependencies become smoother (see Fig.\ref{fig:ff5}). 
Such lowered sensitivity to $Q$ arises due to very restrictive 
constraints (constructed for high $n_0$) corresponding to the 
limited number of parameters with lowered sensitivity to $Q$.
In Fig.\ref{fig:ff6} we see how the invasiveness of the MM and JCs varies with $Q$.  
The characteristics $Df_{min}(Q)$ shows V-shaped ($n_0$-dependent) minima 
that reflect the compensatory effect 
of $f_T$ and $f_J$ localized slightly 
above the threshold values $Q_{thr}(n_0)$.

For further details, we refer the interested reader to Appendix, 
which contains discussion of the periodic extensions of the solutions 
obtained as well as the stability of the RK4 method 
and its comparison with the implicit integration 
methods, regarding specific aspects of our application.

The results of constrained optimization with $n_0=0$ are presented 
in Tab.~\ref{tab:n0eq0}. Similarly, the systematic analysis 
covers $n_0=1$ (see Tab.~\ref{tab:n0eq1}),  $n_0=2$ (Tab.~\ref{tab:n0eq2}), 
$n_0=3$ (Tab.~\ref{tab:n0eq3}), $n_0=4$ (Tab.~\ref{tab:n0eq4}).  
This demonstrates that the optimized components of ${\bf s}_{min}$ 
are interdependent.
In fact, these outputs, apart from underlying complexity
of the responses, answer the questions about the
parametric sensitivity in the particular regions of $(Q, n_0)$ plane. 
Apparently, the components of ${\bf s}_{min}$ differ not
only in the persistence of the prevailing 
discrete values, but in the tendencies to 
make jumps along the $Q$ coordinate as well.
The complexity of the ${\bf s}_{min}(Q, n_0)$ emerges from the 
nonlinearity which destabilizes the form of exogenous pulses.
As the available options for the exogenous influence narrow down
with $n_0$, the relative influence of certain parameters becomes
more apparent.

Although a detailed interpretation 
of the structure of ${\bf s}_{min}$ strings may be difficult, 
a few instructive facts outline their implicit logic.
Some of the results are relatively robust to the specific 
assumptions, including the choice of  $Q$ or $n_0$. 
To illustrate this, we reconsider the solution
${\bf s}_{min}=\left[ 3, -1, -1,+1,-1,\right.$  
$\left. +1,-1\right]$ obtained for $n_0=0$, 
$Q\in \langle 0.01, 0.27\rangle$ (Tab.~\ref{tab:n0eq0}, Tab.~\ref{tab:n0eq1}).
This string shows essential overlap with that of the neighboring interval 
$Q\in \langle 0.27,0.63 \rangle$. 
The only difference is in the sensitivity of the 
third component $(s_{g_{CC}}=+1) \rightarrow (s_{g_{CC}}=-1)$.  
Interestingly, the same component destabilizes 
$[(s_{g_{CC}}=-1) \rightarrow (s_{g_{CC}}=0)]$ when the constraint 
$n_0=1$ is applied. In addition, we see that the therapy 
applied to the both feedback components, 
$s_{h_{CT}}=-1$ and $s_{h_{BT}}=-1$, acts against the tumor influence on OCs and OBs.
This is in line with autocrine inhibition of OCs by $s_{g_{CC}}=-1$ 
which may reduce replication rate of the tumoral cells 
$\sim \alpha_T C^{g_{TC}} T^{g_{TT}}$.
More universal perspective can be captured from 
the comparison in Tabs. \ref{tab:n0eq0}-\ref{tab:n0eq3}. 
Here, the highest stability is achieved 
for $s_{h_{BT}}=-1$ and $s_{g_{BC}}=+1$, while
$s_{g_{CB}}$ is very sensitive to $Q$. 
However, as shown in Tabs. \ref{tab:n0eq4}-\ref{tab:n0eq5},
the role of $s_{g_{CB}}$ is unique, as its nonzero values 
are robust against selection imposed by the constraints 
with $n_0\geq 4$. 
In this case, the zero values of $s_{h_{CT}}$, $s_{g_{CC}}$ $s_{h_{BT}}$, $s_{g_{BB}}$ are quite persistent 
with respect to $Q$.  Within the resulting optimal  
sparse solution, the effect of therapeutically 
relevant timing is controlled by $s_I$ 
coupled with $s_{g_{CB}}=+1$ and $s_{g_{BC}}=-1$.

\begin{table}[bth]
\begin{center}
\begin{tabular}{|r|c||c|cccccc|}
\hline 
\shortstack{
{\tiny inter. length} \\ {\scriptsize $\times$ 100}}  & 
\shortstack{$Q$\\ {\scriptsize interval}}    &   
$ s_I $ &  $ s_{h_{CT}}$  
& $ s_{g_{CC}}$    &   
$ s_{g_{CB}} $      &   $ s_{h_{BT}} $  
&  $ s_{g_{BC}} $  &  $ s_{g_{BB}} $ 
\\
\hline
26 &   < 0.01,  0.27 >   & 3 & {\bf\large -} & {\bf\large -} & {\bf\large +}  & {\bf\large -} & {\bf\large +} & {\bf\large -} \\
36 &   < 0.27,  0.63 >    & 3 & {\bf\large -} & {\bf\large +}  & {\bf\large +}  & {\bf\large -} & {\bf\large +} & 
{\bf\large -} \\
12 &   < 0.63,   0.75 >    & 0 & {\bf\large -} & {\bf\large +}  & {\bf\large -} & {\bf\large -} & {\bf\large +} & {\bf\large -} \\
8  &    < 0.75,   0.83 >    & 1 & {\bf\large -} & {\bf\large +}  & {\bf\large -} & {\bf\large -} & {\bf\large +} & {\bf\large -} \\
42 &   < 0.83,   1.25 >    & 0 & {\bf\large +}  & {\bf\large -} & {\bf\large -} & {\bf\large -} & {\bf\large +} & {\bf\large -} \\
3  &     < 1.25,   1.28 >    & 1 & {\bf\large +}  & {\bf\large -} & {\bf\large -} & {\bf\large -} & {\bf\large +} & {\bf\large -} \\
58 &   < 1.28,   1.86 >    & 0 & {\bf\large -} & {\bf\large -} & {\bf\large -} & {\bf\large -} & {\bf\large +} & {\bf\large -} \\
13 &   < 1.86,   1.99 >    & 1 & {\bf\large -} & {\bf\large -} & {\bf\large -} & {\bf\large -} & {\bf\large +} & {\bf\large -} \\
30  &  < 1.99,   2.29 >    & 3 & {\bf\large -} & {\bf\large +}  & {\bf\large +}  & {\bf\large -} & {\bf\large +} & {\bf\large -} \\ 
102 & < 2.29,   3.31 >   & 1 & {\bf\large -} & {\bf\large +}  & {\bf\large +}  & {\bf\large -} & {\bf\large -} &  {\bf\large -} \\ 
168 & < 3.31,  4.99 >  & 1 & {\bf\large -} & {\bf\large -} & {\bf\large +}  & {\bf\large -} & {\bf\large -} & {\bf\large -} \\
\hline 
\end{tabular}
\caption{The results of constrained optimization with $n_0=0$. 
The relations between $Q$ and ${\bf s}_{min}$ remain persistent for given 
intervals of $Q$ (i.e. for the respective line of the table). 
The lengths of the $Q$ intervals are given in the first column. 
The columns $s_I$, $s_{h_{CT}}, \ldots, s_{g_{BB}}$
correspond to the optimum ${\bf s}_{min}$.  The notation ${\bf\large -}\equiv -1$, $ {\bf\large +}\equiv 1$ 
is used here as well as for all the tables with the similar structure.  
\label{tab:n0eq0}}
\end{center} 
\end{table}

\begin {table}[bth]
\begin{center}
\begin{tabular}{|r|c||c|cccccc|}
\hline 
\shortstack{
{\tiny inter. length} \\ {\scriptsize $\times$ 100}}  & 
\shortstack{$Q$\\ {\scriptsize interval}}    &   
$ s_I             $   &  
$ s_{h_{CT}}$   & 
$ s_{g_{CC}}$   &   
$ s_{g_{CB}} $  &   
$ s_{h_{BT}} $   &  
$ s_{g_{BC}} $   &  
$ s_{g_{BB}} $ 
\\
\hline
25 &  <0.01, 0.26> & 3 & {\bf\large -} & {\bf\large -} & {\bf\large +} & {\bf\large -} & {\bf\large +} & 0  \\
<  1 &  <0.27, 0.27> & 3 & {\bf\large -} & 0  & {\bf\large +} & {\bf\large -} & {\bf\large +} & {\bf\large -} \\
35 & <0.28, 0.63> & 3 & {\bf\large -} & {\bf\large +} & {\bf\large +} & {\bf\large -} & {\bf\large +} & 0  \\
6 & <0.64, 0.70> & 0 & {\bf\large -} & {\bf\large +} & {\bf\large -} & {\bf\large -} & {\bf\large +} & 0  \\
14 & <0.71, 0.85> & 0 & {\bf\large +} & 0 & {\bf\large -} & {\bf\large -} & {\bf\large +} & {\bf\large -} \\
20 & <0.86, 1.06> & 0 & {\bf\large -} & 0 & {\bf\large -} & {\bf\large -} & {\bf\large +} & {\bf\large -} \\
8 &  <1.07, 1.15> & 0 & {\bf\large +} & {\bf\large -} & {\bf\large -} & {\bf\large -} & {\bf\large +} & 0  \\
18 & <1.16, 1.34> & 0 & 0 & {\bf\large -} & {\bf\large -} & {\bf\large -} & {\bf\large +} & {\bf\large -} \\
28 & <1.35, 1.63> & 0 & {\bf\large -} & {\bf\large -} & {\bf\large -} & {\bf\large -} & {\bf\large +} & 0 \\
92 & <1.64, 2.56> & 0 & {\bf\large +} & {\bf\large +} & 0 & {\bf\large -} & {\bf\large +} & {\bf\large -} \\
44 & <2.57, 3.01> & 1 & {\bf\large +} & {\bf\large +} & 0 & {\bf\large -} & {\bf\large +} & {\bf\large -} \\
80 & <3.02, 3.82> & 1 & {\bf\large -} & 0 & {\bf\large +} & {\bf\large -} & {\bf\large -} & {\bf\large -} \\
19 & <3.83, 4.02> & 1 & {\bf\large -} & {\bf\large -} & {\bf\large +} & {\bf\large -} & {\bf\large -} & 0 \\
< 1 & <4.03, 4.03> & 1 & {\bf\large -} & {\bf\large -} & 0 & {\bf\large -} & {\bf\large -} & {\bf\large -} \\
95 & <4.04, 4.99> & 1 & {\bf\large -} & {\bf\large -} & {\bf\large +} & {\bf\large -} & {\bf\large -} & 0 \\
\hline 
\end{tabular}
\caption{The results of the $f(t_S,t_E,{\bf s})$ minimization constrained by $n_0=1$. 
The differences in ${\bf s}_{min}$ obtained for different intervals of $Q$. 
Total persistence of the $s_{h_{BT}}=-1$. Relatively high persistence of 
$s_{g_{BC}}=1$ for $Q<3$ (See results of synergistic analysis in sec.\ref{sec:Synerg} below). 
Highest number of zeros and highest frequency of the 
changes between $-1$ and $0$ 
is shown in the column $s_{g_{BB}}$. 
\label{tab:n0eq1}}
\end{center}
\end{table}

\begin {table}[bth]
\begin{center}
\begin{tabular}{|r|c|c|cccccc|}
\hline 
\shortstack{ {\tiny inter. length} \\ {\scriptsize $\times$ 100}}  & 
\shortstack{$Q$\\ {\scriptsize interval}}    &   
$ s_I $ &  
$ s_{h_{CT}}$  
& $ s_{g_{CC}}$      &   
$ s_{g_{CB}} $        &    $ s_{h_{BT}} $  
&  $ s_{g_{BC}} $    &   $ s_{g_{BB}} $ 
\\
\hline
  4  & <0.01, 0.05>   & 3  & 0  & {\bf\large -}  &{\bf\large +}  & {\bf\large -} &{\bf\large +}  & 0 \\
 42  & <0.06, 0.48>   & 3  & {\bf\large -}  & 0  &{\bf\large +}  & {\bf\large -} &{\bf\large +}  & 0 \\
 13  & <0.49, 0.62>   & 3  & 0  &{\bf\large +}  &{\bf\large +}  & {\bf\large -} &{\bf\large +}  & 0 \\
 15  & <0.63, 0.78>   & 0  &{\bf\large +}  & 0  & {\bf\large -}  & {\bf\large -} &{\bf\large +}  & 0 \\
 10  & <0.79, 0.89>   & 0  & 0  & 0  & {\bf\large -}  & {\bf\large -} &{\bf\large +}  & {\bf\large -} \\
 17  & <0.90, 1.07>   & 0  & {\bf\large -}  & 0  & {\bf\large -}  & {\bf\large -} &{\bf\large +}  & 0 \\
  2  & <1.08, 1.10>   & 1  & {\bf\large -}  & 0  & {\bf\large -}  & {\bf\large -} &{\bf\large +}  & 0 \\
 23  & <1.11, 1.34>   & 0  & 0  & {\bf\large -}  & {\bf\large -}  & {\bf\large -} &{\bf\large +}  & 0 \\
 12  & <1.35, 1.47>   & 1  & 0  & {\bf\large -}  & {\bf\large -}  & {\bf\large -} &{\bf\large +}  & 0 \\
108  & <1.48, 2.56>   & 0  &{\bf\large +}  &{\bf\large +}  & 0  & {\bf\large -} &{\bf\large +}  & 0 \\
 44  & <2.57, 3.01>   & 1  &{\bf\large +}  &{\bf\large +}  & 0  & {\bf\large -} &{\bf\large +}  & 0 \\
100  & <3.02, 4.02>   & 1  & {\bf\large -}  & 0  &{\bf\large +}  & {\bf\large -} & {\bf\large -}  & 0 \\
  0  & <4.03, 4.03>   & 1  & {\bf\large -}  & {\bf\large -}  & 0  & {\bf\large -} & {\bf\large -}  & 0 \\
  95 & <4.04, 4.99>   & 1  & {\bf\large -}  & 0  &{\bf\large +}  & {\bf\large -} & {\bf\large -}  & 0 \\
\hline 
\end{tabular}
\caption{The optimized components of ${\bf s}_{min}$ 
(see Eq.(\ref{eq:smin}))
obtained for the constraint 
$n_0=2$ (see Eq.(\ref{eq:n0})). 
\label{tab:n0eq2}}
\end{center} 
\end{table}

\begin {table}[bth]
\begin{center}
\begin{tabular}{|r|c|c|cccccc|}
\hline 
\shortstack{ {\tiny inter. length} \\ {\scriptsize $\times$ 100}}  & 
\shortstack{$Q$\\ {\scriptsize interval}}    &   
$ s_I $ &  
$ s_{h_{CT}}$  
& $ s_{g_{CC}}$      &   
$ s_{g_{CB}} $        &    $ s_{h_{BT}} $  
&  $ s_{g_{BC}} $    &   $ s_{g_{BB}} $ 
\\
\hline
  <1   &  <0.01, 0.01>   &  3 & 0 & {\bf\large -}         & {\bf\large +} & {\bf\large -} & 0 & 0 \\
 59  &  <0.02, 0.61>   &  3 & 0 & 0  &{\bf\large +}  & {\bf\large -} &{\bf\large +}  & 0 \\
 31  & <0.62, 0.93>    &  0 & 0 & 0  & {\bf\large -}  & {\bf\large -} &{\bf\large +}  & 0 \\
 25  & <0.94, 1.19>    &  1 & 0 & 0  & {\bf\large -}  & {\bf\large -} &{\bf\large +}  & 0 \\
 26  & <1.20, 1.46>    &  3 & 0 & 0  &{\bf\large +}  & {\bf\large -} &{\bf\large +}  & 0 \\
141 & <1.47, 2.88>    &  0 & 0 & {\bf\large +}        & 0                 & {\bf\large -}  & {\bf\large +} & 0 \\
210 & <2.89, 4.99>    &  1 & 0 & 0  &{\bf\large +}  & {\bf\large -} & {\bf\large -}  & 0 \\
\hline       
\end{tabular}
\caption{The effect of the constraint $n_0=3$.
\label{tab:n0eq3}} 
\end{center}
\end{table}

\begin {table}[bth]
\begin{center}
\begin{tabular}{|r|c||c|cccccc|}
\hline 
\shortstack{ {\tiny inter. length} \\ {\scriptsize $\times$ 100}}  & 
\shortstack{$Q$\\ {\scriptsize interval}}    &   
$ s_I $ &  
$ s_{h_{CT}}$  
& $ s_{g_{CC}}$      &   
$ s_{g_{CB}} $        &    $ s_{h_{BT}} $  
&  $ s_{g_{BC}} $    &   $ s_{g_{BB}} $ 
\\
\hline
169 & <0.01, 1.70>  & 3 & 0 & 0 & {\bf\large +}& {\bf\large -} & 0 & 0 \\
35   & <1.71, 2.06>  & 0 & 0 & {\bf\large +}& 0 & 0 & {\bf\large +}& 0 \\ 
44   & <2.07, 2.52>  & 1 & 0 & 0 & {\bf\large +}& 0 & {\bf\large -} & 0 \\ 
247 & <2.53, 4.99>  & 1 & 0 & 0 & {\bf\large +}& 0 & {\bf\large -} & 0 \\
\hline       
\end{tabular}
\caption{The optimized ${\bf s}_{min}$ obtained for $n_0=4$. 
Compared to the low $n_0$, longer intervals in $Q$ occur.
\label{tab:n0eq4}}
\end{center}
\end{table}

\begin{table}[bth]
\begin{center}
\begin{tabular}{|r|c||c|cccccc|}
\hline 
\shortstack{ {\tiny inter. length} \\ {\scriptsize $\times$ 100}}  & 
\shortstack{$Q$\\ {\scriptsize interval}}    &   
$  s_I $ &  
$  s_{h_{CT}}$  
& $ s_{g_{CC}}$      &   
$     s_{g_{CB}} $        &    $ s_{h_{BT}} $  
&  $ s_{g_{BC}} $    &   $ s_{g_{BB}} $ 
\\
\hline
498 & <0.01,4.99> & 3 & 0 & 0 & {\bf\large +} & 0 & 0 & 0 
\\
\hline 
\end{tabular}
\caption{The optimization result for $n_0=5$ emphasizing the key role of 
$s_{g_{CB}}$.
\label{tab:n0eq5}}
\end{center}
\end{table}

\section{Alternative scalarizations: linkage between the worst-case and the best-case}\label{sec:alters}

\begin{figure}[bth]
\includegraphics[width=0.98\linewidth]{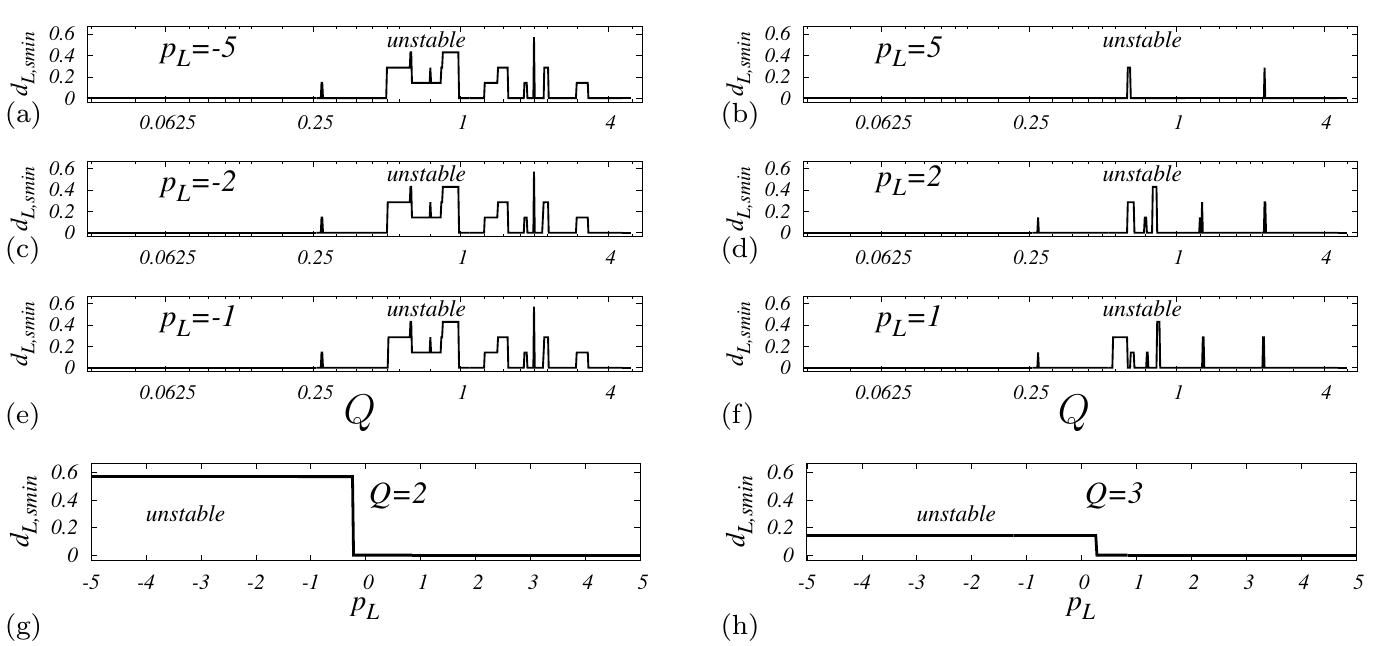}
\caption{ $Q$ and $p_L$ dependencies of $d_{L,smin}$ distances introduced
to analyze relative stability of the optima 
of the alternative scalarizations discussed in the section \ref{sec:alters}. 
The panels (b), (d), (f), (g), (h) show that 
increasing positive values of $p_L$ results in gradual 
vanishing of instability,  in agreement
with the expectation (we note that the worst-case 
scenario corresponds to $p_L\rightarrow \infty$). 
The distances calculated for the negative 
$p_L$ values $-1,-2,-5$ [panels (a) ,(c), (e)] are nearly identical to each other. 
Specifically, the choice of small $Q$ can guarantee higher stability,
thus lowering sensitivity to the way in which the 
scalarization is performed. Calculated for the constraint $n_0=0$.}
\label{fig:ffX8}
\end{figure}

In this section we address the particular issue how the optimization output depends
on the particular scalarization of the original multi-objective problem considering
the objectives $f_T$ and $f_J$, which further opens the question
of the stability of the optima. For a stable scalarization, the small variation 
in the original formula results in the small change in the optimum obtained. 
To investigate the issue of eventual alternatives
to the benchmark scalarization by the worst-case scenario, 
we propose the following framework requirements: 
(a)~the alternative scalarization is parametrized by the single real parameter 
which quantifies deviation from the worst-case limit; 
(b)~the invariance under the exchange of $f_T$ and $f_J$;
(c)~scalarizations proceed from the generalized mean
that expresses the central tendencies of $f_T$ and $f_J$. 
The alternative which fulfills the above requirements is
{\em Lehmer mean} \cite{Beliakov2015}
\begin{equation}
f_{L}(p_L)  =  \frac{f_T^{p_{L}} + f_J^{p_{L}}}{f_T^{p_{L}-1} + f_J^{p_{L}-1}}\, 
\end{equation}
which we use  in the further analysis.
The family of scalarized variants $\{f_{L}(p_L)\,|$ $\, p_L\in \mathbb{R}\}$ 
uses $p_L$ to bridge the worst-case $f_{L}(p_L\rightarrow \,\infty)=
\max \{f_T, f_J\}$
and the best-case $f_{L}(p_L\rightarrow \,-\infty)= \min \{f_T, f_J\}$
(therapeutically infeasible) limits. 
To quantify the level of stability, the 
global optimum
${\bf s}_{L,min}=\arg$\, $\min_{s\in \Omega} f_L(p_L) $ is compared  
with the optimum corresponding to the worst-case limit  
via the alternative mean 
Hamming distance
\begin{equation}
d_{L,smin}(p_L) = \frac{1}{7}\, 
\sum_{\,\,\,\,\,\,\,\{\forall \, {\bf s} \,\, \mbox{\tiny components}\}} 
\mbox{\Large ${\mathds{1}}$}
\left(
\arg \min_{s\in \Omega} f_L(p_L) 
\neq 
\arg \min_{s\in \Omega} f\,
\right)\,.  
\end{equation}
constructed in complete analogy to Eq.(\ref{eq:dH}).
The results of the numerical analysis of $d_{L,smin}$
corresponding to the case $n_0=0$ (other constraints
show an analogous behavior) are presented in Fig.\ref{fig:ffX8}.
They confirm qualitative differences between negative (unstable, 
i.e. considerably different from the worst-case), 
and stable domains of sufficiently 
large $p_L>0$. The additional information obtained from the 
calculations is that scalarizations performed for the
small $Q$ region (approximately $Q \lesssim 0.2$) are relatively well stabilized.
This contrasts with the optimal ${\bf s}_{L,min}$ 
corresponding to the higher $Q$ values,
where sensitivity to $p_L$ becomes pronounced. 
We see that computational (and, eventually, therapeutic) 
problems lie mainly in the domain of high $Q$ values,
where the irregular and less persistent behavior of $d_{L,smin}(p_L)$ can be observed.

\section{Modified objective function with the penalty term, sparse optimization}\label{se:lamPen}

We completed our work by experimenting with the alternative worst-case scalarized 
objective function
\begin{equation} 
f(t_S, t_E, {\bf s},\lambda)   =  
\max \{\, f_T(t_S, t_E, {\bf s}), \, 
f_J(t_S, t_E, {\bf s}) \}  + 
\lambda Q \left(1- 
\frac{n_0({\bf s})}{6}\right)\,, 
\label{eq:fpen} 
\end{equation}
supplemented by the extra {\em penalty term} 
$\lambda Q (1- $  $ n_0({\bf s})/6)$ 
aimed at the selection of representative parameters \cite{Lethi2015}.
The side effects of the therapy are proportional to the regularization 
parameter $\lambda\geq 0$ and previously introduced strength of the therapeutic action $Q$.  
The factor $ (1 - n_0({\bf s})/6) \in (0,1)$ accounts for the fraction 
of nonzero ${\bf s}$ components.
To sum up, comparing it to 
the function $f(t_S,t_E,{\bf s})$ Eq.(\ref{eq:maxf}), the above 
formulation minimizes unpredictable toxic side effects of the 
therapy \cite{Foo2009} by 
preferring weaker interventions. Moreover, 
strict constraint on ${\bf s}$ no longer applies.
In addition, we note that the sparsity-inducing penalization 
introduced by Eq.(\ref{eq:fpen})
is analogous to that adopted when considering the effective 
Markowitz portfolio diversification \cite{Brodie2009}.

\begin{figure}[bth]
\mbox{\begin{tabular}{l}
\includegraphics[width=0.82\linewidth]{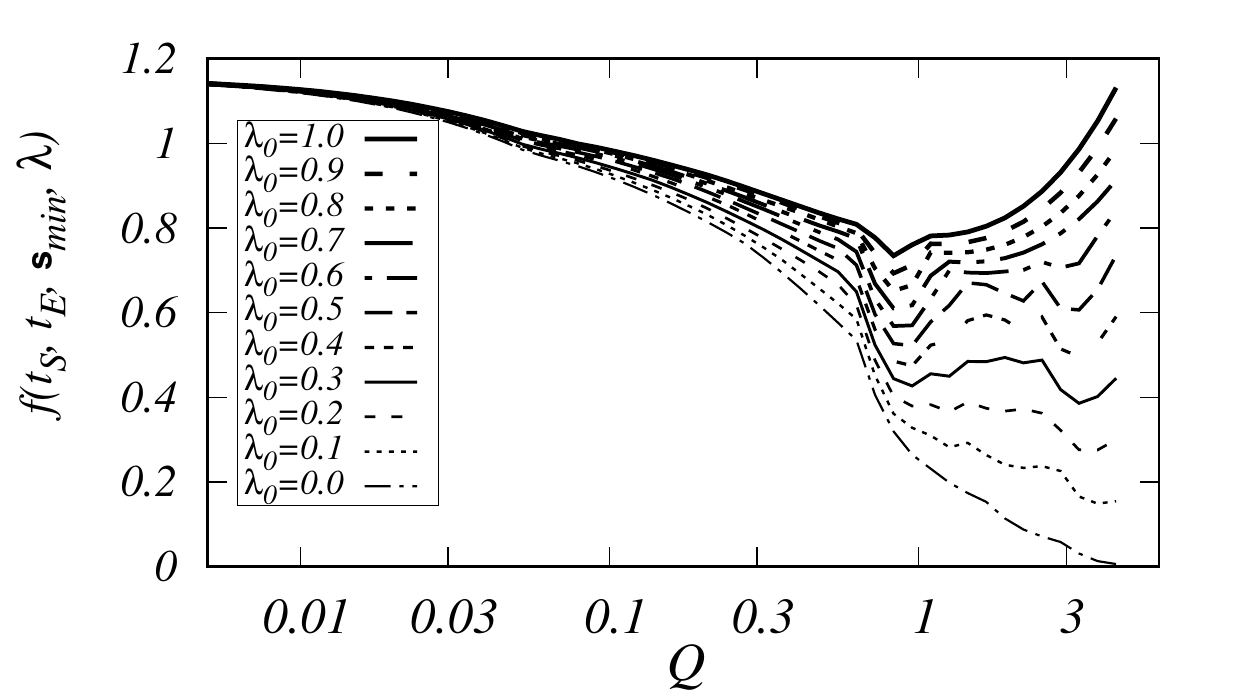}
\\
\vspace*{-12mm}
\\
\mbox{\large (a)} 
\\
\vspace*{-2mm}
\\ 
\hspace*{11.5mm} \includegraphics[width=0.78\linewidth]{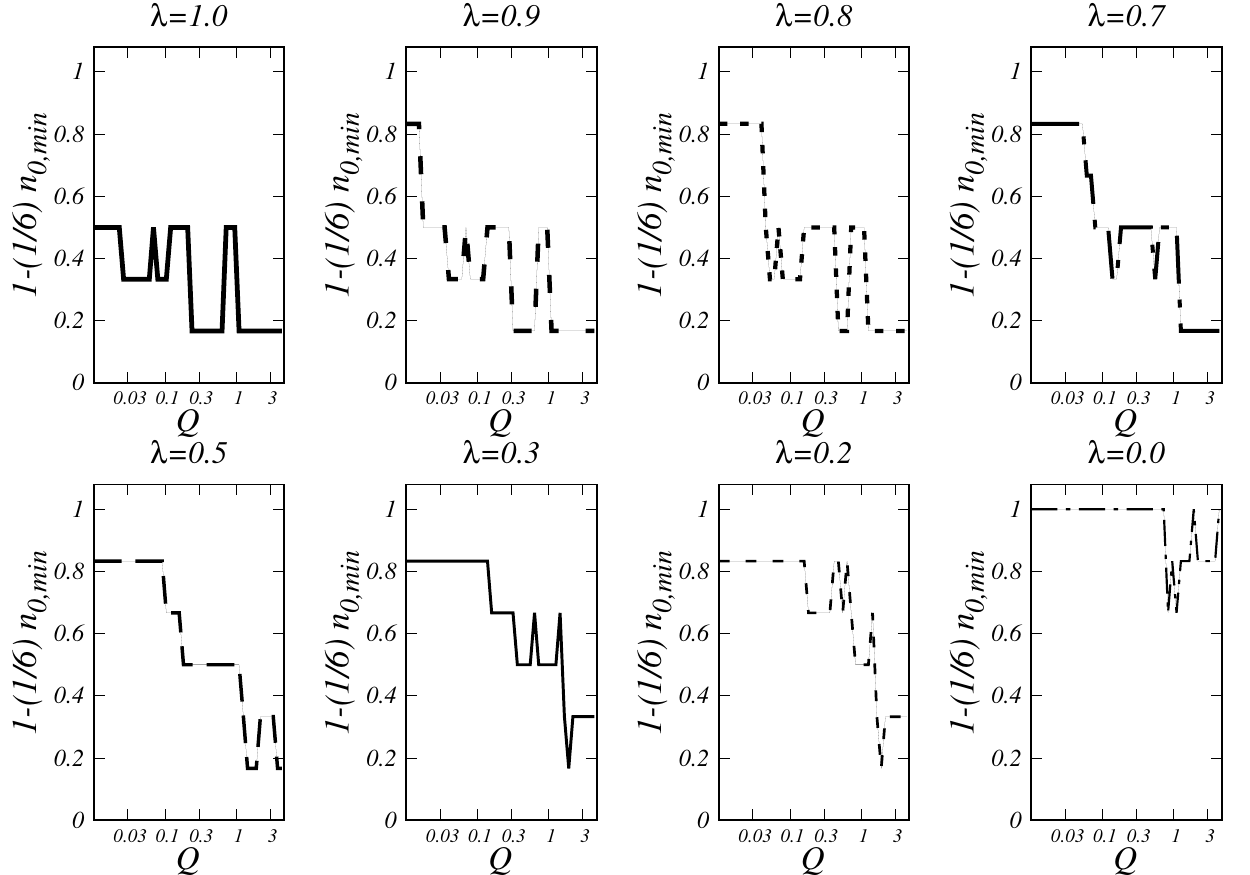}
\\
\vspace*{-12mm}
\\
\mbox{\large (b)} 
\end{tabular}}
\caption{The $Q$-dependence of the optimized  
alternative objective function values 
$f(t_S, t_E, {\bf s}_{min}, \lambda)$ 
[see Eq.(\ref{eq:fpen})] calculated for 
$\lambda=0.0, \,0.1,\,\ldots, 1.0 $ 
[panel (a)] supplemented with multiple plots 
of $1-(1/6) (n_0)_{min}$ quantity [panel (b)]. 
The plots demonstrate 
how the fraction of non-zero ${\bf s}_{min}$ components varies with $Q$.}
\label{fig:ff9}
\end{figure}

In Fig.\ref{fig:ff9} we present numerical results obtained 
for the function Eq.(\ref{eq:fpen}). 
They reveal that the optimum of $f(t_S, t_E, {\bf s}, \lambda)$
exists not only in ${\bf s}$, but along $Q$ for 
large enough $\lambda$ ($\lambda>0.3$)
as well. Highly inefficient effects correspond to $Q\gtrsim 0.8 $. 
As expected [Fig.\ref{fig:ff9}(b)], 
in all the studied cases increase in 
$Q$ reduces the number of non-zero 
elements proportional to $1-  (1/6)(n_0) _ {min}$
where $(n_0)_{min}=n_0({\bf s}_{min})$. 

\vspace*{1mm}

\section{The case of asymmetric pulses}\label{sec:APulse}  

The symmetric pulses [see Eq.(\ref{eq:sigmt})] have been 
originally designed as a tool of the sensitivity analysis with unclear relationship to pharmacological 
applications. Because the pharmacologic characteristics of the pulse 
protocols \cite{Ananthakrishnan2010} exhibit, in general, non symmetric, right tailed shape in time, 
we introduce more realistic asymmetric model in this section.
By focusing on the exogenous changes, the model is introduced in two main steps. 
The first is represented by the auxiliary formula
\begin{equation}
\phi_a(t',\sigma_a,W_a,\tau_a) = 
\exp\left[ -\frac{(t')^2}{2\sigma_a^2} 
\left( 1-{\rm logist}\left(\frac{t'}{W_a}\right)\right) - \frac{t'}{\tau_a} 
{\rm logist}\left(\frac{t'}{W_a}\right)\right]
\label{eq:phiA1} 
\end{equation}
incorporating the logistic function ${\rm logist}(\zeta)=1/(1+\exp(-\zeta)) 
\in (0,1)$  of the dimensionless argument $\zeta=t'/W_a$, 
rescaling the time argument by the transition interval width $W_a$. 
The logistic local in time weighting causes 
that the characteristic time scale of the infusion/absorption regime 
$1/\sigma_a$ is smoothly changing (on the time scale $1/W_a$) to the 
late-time exponential drug elimination/excretion 
(depending on the metabolic conditions) with the characteristic 
time $\tau_a$. 

\begin{figure}[bth]
\includegraphics[angle=-90,width=0.8\linewidth]{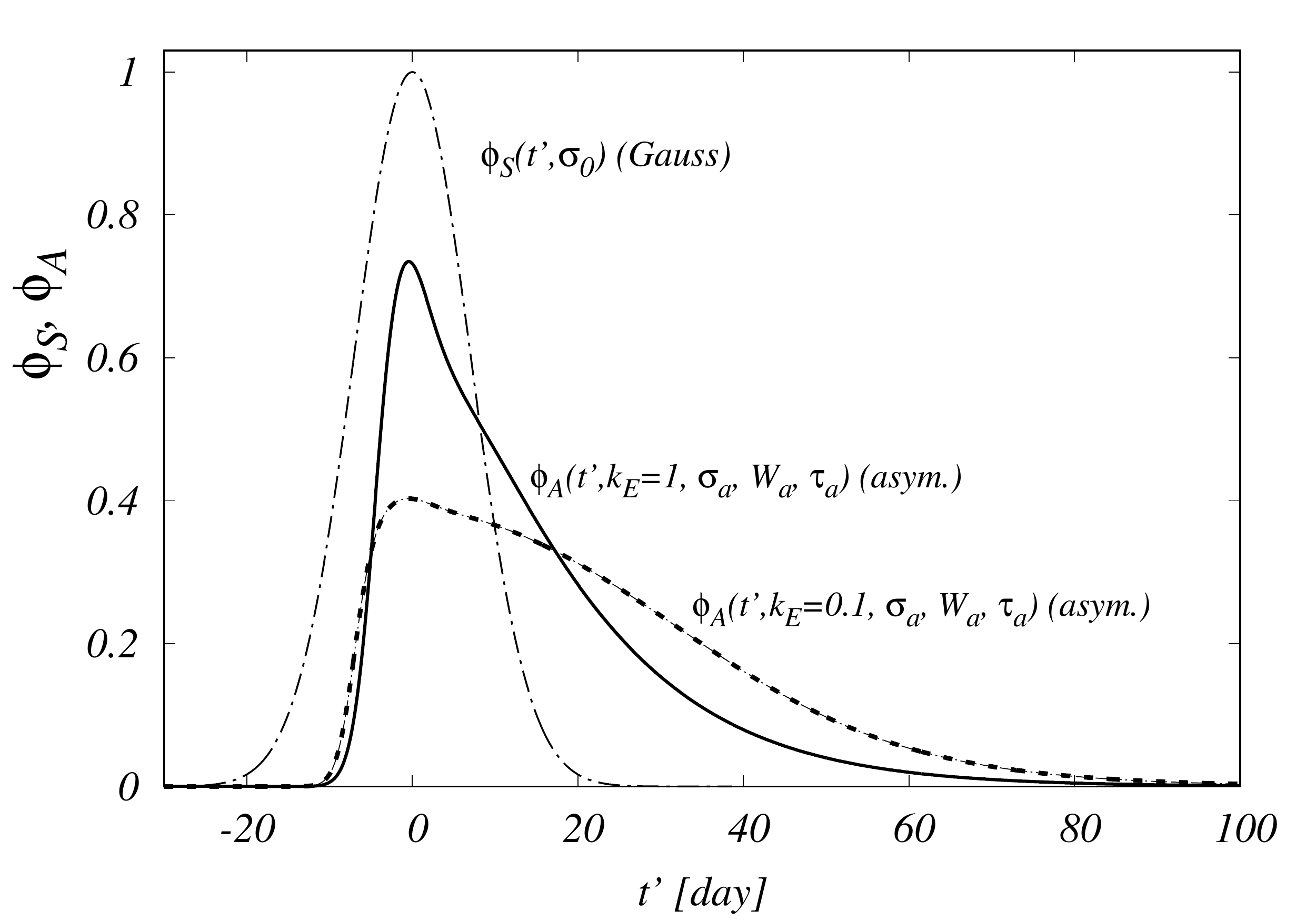}
\caption{The figure illustrates the model of the asymmetric pulses given by 
Eqs.(\ref{eq:phiA1}) and (\ref{eq:phiA2}) which replace symmetric form 
[Eq.(\ref{eq:sigmt})]. The asymmetric pulses have been used 
in the numerical optimizations.}
\label{fig:phA1A2}
\end{figure}

The second step of the pulse definition reflects the fact that the impact
of the drug depends not only on its concentration but on the effectiveness
of the binding on the respective receptor as well, and it may be  
described by the transformation   
\begin{equation}
\phi_A(t') \equiv \phi_A(t',k_E,\sigma_a, W_a,\tau_a) =  
n_E \frac{\phi_a(t',\sigma_a, W_a,\tau_a)}
{k_E + \phi_a(t',\sigma_a, W_a,\tau_a)}\,
\label{eq:phiA2} 
\end{equation}
corresponding to the sigmoid model \cite{Ananthakrishnan2010}. The form Eq.(\ref{eq:phiA2}) 
replaces Eq.(\ref{eq:sigmt}); $n_E$ is the normalization parameter 
derived from the condition $\int_{-\infty}^{\infty}{\rm d}t\, $ $
\phi_S(t,\sigma_0) $ $=\int_{-\infty}^{\infty} {\rm d}t $ $
\phi_A(t,.)$. The normalization is proposed
to achieve equivalence between the Gaussian 
($\sigma_0 = 7 day$) and the asymmetric pulses defined 
by $k_E,$ $\sigma_a,$ $W_a,$ $\tau_a$. 
In the normalized case shorter duration is accompanied with 
enhanced intensity and vice versa. Two variants of the parameter $k_E \in \{0.1, 1\}$ 
controlling the drug efficiency have been used in the numerical calculations: 
(a) $k_E=1$ corresponding to the high efficiency regime 
with normalization prefactor $n_E=1.46355$; 
(b) $k_E=0.1$ as a model of low drug efficiency with $n_E=0.44288$. 
In Fig.\ref{fig:phA1A2}, the shape of asymmetric pulses is calculated for the 
numerical parameters $W_a=2\, day,$ $\sigma_a=3 \,day,$ $\tau_a= 14\, day$
that we used in further optimizations.

The detailed comparison of the optimization results obtained for 
the symmetric and asymmetric pulses for the respective penalty 
functions [Eq.(\ref{eq:fpen})] with the regularization parameters 
$\lambda\in \{0,0.2,0.4,0.6,0.8,1\}$ is presented in Tab.~\ref{tab:asymversG}. 
In general, we conclude that (a) the pulse shape affects the optimization result,
nevertheless the correlation of the optimal strings corresponding to the symmetric and asymmetric pulses is positive; 
(b) the values of $s_I$ encoding the pulse width are more susceptible to 
    the pulse symmetry/asymmetry and drug efficiency ($k_E$);  
(c) the nonzero value of $s_{g_{BB}}$ may disrupt the optimization results;  
(d) the value of $s_{g_{BC}}$ is relatively persistent 
for different $\lambda$ which coincides with its integral synergistic effect 
identified in sec.\ref{sec:Synerg}.

\vspace*{-1mm}

{\tiny 
\begin{longtable}{| l | l l | l | l l l l l l l |l|}
\multicolumn{4}{l} 
{\LARGE (A)\,\,\, \fbox{$\lambda=0$}} 
\\
\hline
   {\large Pulse}  & {\large $k_E$} & {\large $Q$}  &  {\large $f_{min}(\lambda)$}   & 
            {\large $s_I$} & {\large $s_{h_{CT}}$} & {\large $s_{g_{CC}}$}  & 
            {\large $s_{g_{CB}}$} & {\large $s_{h_{BT}}$}    & 
            {\large $s_{g_{BC}}$} & {\large $s_{g_{BB}}$}    & 
            {\large $d_{H,A-G}$}  \\
\hline
asym.L & 0.1  & 0.2 & 0.7898  & 2$_{\Conv}$ & {\bf \large -}  & {\bf \large -} & {\bf \large +} & {\bf \large -} 
&  {\bf \large +} & {\bf \large -} & 1 \\
asym.H & 1.0  & 0.2 & 0.7792  & 2$_{\Conv}$ & {\bf \large -}  & {\bf \large -} & {\bf \large +} & {\bf \large -} &  1 & {\bf \large -} & 1 \\ 
Gauss  &      & 0.2 & 0.8200  & 3  & {\bf \large -}  & {\bf \large -} & {\bf \large +} & {\bf \large -} &  
{\bf \large +} & {\bf \large -} &   \\ 
\hline
asym.L & 0.1     &  0.4 & 0.6246 & 2$_{\Conv}$ & {\bf \large -}  & {\bf \large -}$_{\Conv\bullet}$ & 
{\bf \large +}   & {\bf \large -} & {\bf \large +} & 
{\bf \large -}   & 2 (1)\\ 
asym.H & 1.0     &  0.4 & 0.6105 & 2$_{\Conv}$ & {\bf \large -}  &  
{\bf \large +}   & {\bf \large +} & {\bf \large -} & {\bf \large +} & {\bf \large -} & 
1 \\ 
Gauss  &      &  0.4 & 0.6667 & 3  & {\bf \large -}  &  {\bf \large +}   
& {\bf \large +} & {\bf \large -} & {\bf \large +} & {\bf \large -} & \\ 
\hline 
asym.L & 0.1 &  0.6 & 0.4892  &  2$_{\Conv}$ & {\bf \large -}  & {\bf \large +} & {\bf \large +} & {\bf \large -} & {\bf \large +} & {\bf \large -} & 1 \\
asym.H & 1.0 &  0.6 & 0.4637  &  2$_{\Conv}$ & {\bf \large -}  & {\bf \large +} & {\bf \large +} & {\bf \large -} & {\bf \large +} & {\bf \large -} & 1 \\
Gauss  &     &  0.6 & 0.5492  &  3  & {\bf \large -}  & {\bf \large +} & {\bf \large +} & {\bf \large -} & {\bf \large +} & {\bf \large -} & \\
\hline 
asym.L &0.1 &  0.8 & 0.3734   & 2$_{\Conv\bullet}$ & {\bf \large -}$_{\Conv}$ & {\bf \large +} &  {\bf \large }$_{\Conv}$ & {\bf \large -} & {\bf \large +} & {\bf \large -} & 3 (1)\\
asym.H &1.0 &  0.8 & 0.3513   & 3$_{\Conv}$  & {\bf \large -}$_{\Conv}$ & {\bf \large +} &  {\bf \large }$_{\Conv}$ & {\bf \large -} & {\bf \large +} & {\bf \large -} & 3 \\
Gauss  &     &  0.8 & 0.3395   & 0   &  0  & 0 & {\bf \large -}  & {\bf \large -} & {\bf \large +} & {\bf \large -} & \\ 
\hline 
asym.L & 0.1 & 1.0 &  0.2812 & 2$_{\Conv\bullet}$ & {\bf \large -} & {\bf \large +}$_{\Conv}$ & {\bf \large +}$_{\Conv}$ & {\bf \large -} & {\bf \large +} & {\bf \large -} & 3 (1) \\
asym.H & 1.0 & 1.0 &  0.2650 & 3$_{\Conv}$  & {\bf \large -} & {\bf \large +}$_{\Conv}$ & {\bf \large +}$_{\Conv}$ & {\bf \large -} & {\bf \large +} & {\bf \large -} & 3 \\
Gauss  &     & 1.0 &  0.2488 & 0   & {\bf \large -} & 0  &{\bf \large -}  & {\bf \large -} & {\bf \large +} & {\bf \large -} &   \\
\hline
\hline
\multicolumn{4}{l} {\LARGE (B) \fbox{$\lambda=0.2$}} 
\\ 
\hline
asym.H  & 0.1 & 0.2 & 0.8231 & 2$_{\Conv}$ & {\bf \large -} & {\bf \large -}$_{\Conv}$   & {\bf \large +} & {\bf \large -} & {\bf \large +} & 0 &   2   \\
asym.L  & 1.0 & 0.2 & 0.8125 & 2$_{\Conv}$ & {\bf \large -} & {\bf \large -}$_{\Conv}$ & {\bf \large +} & {\bf \large -} & {\bf \large +} & 0 &   2   \\ 
Gauss   &     & 0.2 & 0.8512 & 3          & {\bf \large -} &  0             & {\bf \large +} & {\bf \large -} & {\bf \large +} & 0 &       \\
\hline 
asym.L    & 0.1 & 0.4 & 0.6820 & 2$_{\Conv}$ & {\bf \large -} & 0 &  1 & {\bf \large -} & {\bf \large +} & 0 &  1 \\
asym.H    & 1.0 & 0.4 & 0.6662 & 2$_{\Conv}$ & {\bf \large -} & 0 &  1 & {\bf \large -} & {\bf \large +} & 0 &  1 \\
Gauss     &     & 0.4 & 0.7310 & 3           & {\bf \large -} & 0 &  1 & {\bf \large -} & {\bf \large +} & 0 &    \\ 
\hline 
asym.L    & 0.1 & 0.6 & 0.5828 & 2$_{\Conv}$ & {\bf \large -}$_{\Conv}$ & 0$_{\Conv\bullet}$  & {\bf \large +} & {\bf \large -}     & {\bf \large +}   & 0   &  3 (1) \\
asym.H    & 1.0 & 0.6 & 0.5637 & 2$_{\Conv}$ & {\bf \large -}$_{\Conv}$ & {\bf \large +}            & {\bf \large +} & {\bf \large -}     & {\bf \large +}   & 0   &  2 \\
Gauss     &     & 0.6 & 0.6451 & 3           &  0           & {\bf \large +}            & {\bf \large +} & {\bf \large -}     & {\bf \large +}   & 0   &    \\
\hline 
asym.L     & 0.1 & 0.8 & 0.5050 & 2$_{\Conv}$ & {\bf \large -}$_{\Conv\bullet}$ & {\bf \large +}$_{\Conv}$  & {\bf \large +}$_{\Conv}$ & {\bf \large -} & 0$_{\Conv\bullet}$ &  0  & 5 (2) \\ 
asym.H     & 1.0 & 0.8 & 0.4810 & 2$_{\Conv}$ &  0           & {\bf \large +}$_{\Conv} $ & {\bf \large +}$_{\Conv}$ & {\bf \large -} & {\bf \large +}           &  0  & 3 \\ 
Gauss      &     & 0.8 & 0.4195 & 0           &  0           & 0            & {\bf \large -}          & {\bf \large -} & {\bf \large +}           &  0  &   \\
\hline 
asym.L     & 0.1 & 1.0 & 0.4362 & 2$_{\Conv\bullet}$ & 0 & {\bf \large +}$_{\Conv}$ & {\bf \large +}$_{\Conv}$ & {\bf \large -} & 0$_{\Conv\bullet}$ & 0  & 4 (2)  \\
asym.H     & 1.0 & 1.0 & 0.4165 & 3$_{\Conv}$ & 0 & {\bf \large +}$_{\Conv}$ & {\bf \large +}$_{\Conv}$ & {\bf \large -} & {\bf \large +}           & 0  & 3  \\ 
Gauss      &     & 1.0 & 0.3796 & 1           & 0 & 0           &{\bf \large -}           & {\bf \large -} & {\bf \large +}           & 0  &    \\
\hline
\hline
\multicolumn{4}{l} {\LARGE (C) \fbox{$\lambda=0.4$}} 
\\ 
\hline
asym.L       & 0.1 & 0.2 & 0.8560 & 2$_{\Conv}$ & {\bf \large -} & {\bf \large -}$_{\Conv\bullet}$ & {\bf \large +} & {\bf \large -} & 0$_{\Conv\bullet}$ & 0 & 3 (2) \\
asym.H       & 1.0 & 0.2 & 0.8434 & 2$_{\Conv}$ & {\bf \large -} &  0           & {\bf \large +} & {\bf \large -} & {\bf \large +}           & 0 & 1 \\
Gauss        &     & 0.2 & 0.8778 & 3 &           {\bf \large -} &  0           & {\bf \large +} & {\bf \large -} & {\bf \large +}           & 0 &   \\
\hline 
asym.L       & 0.1 & 0.4 & 0.7273 & 2$_{\Conv}$ & {\bf \large -}$_{\Conv\bullet}$ & 0 & {\bf \large +} & {\bf \large -} & 0$_{\Conv\bullet}$ & 0  & 3 (2) \\
asym.H       & 1.0 & 0.4 & 0.7176 & 2$_{\Conv}$ &  0           & 0 & {\bf \large +} & {\bf \large -} & {\bf \large +}           & 0  & 1 \\
Gauss        &     & 0.4 & 0.7744 & 3           &  0           & 0 & {\bf \large +} & {\bf \large -} & {\bf \large +}           & 0  &   \\
\hline
asym.L       & 0.1 & 0.6 & 0.6334 & 2$_{\Conv}$ & 0 & 0 & {\bf \large +} & {\bf \large -} & 0$_{\Conv}$ & 0  & 2 \\
asym.H       & 1.0 & 0.6 & 0.6261 & 2$_{\Conv}$ & 0 & 0 & {\bf \large +} & {\bf \large -} & 0$_{\Conv}$ & 0  & 2 \\
Gauss        &     & 0.6 & 0.7104 & 3           & 0 & 0 & {\bf \large +} & {\bf \large -} & {\bf \large +}           & 0  &   \\ 
\hline 
asym.L       & 0.1 & 0.8 & 0.5624 & 2$_{\Conv\bullet}$ & 0 & 0 & {\bf \large +}$_{\Conv}$ &  0$_{\Conv\bullet}$  & 0$_{\Conv}$ & 0 & 4 (2)\\ 
asym.H       & 1.0 & 0.8 & 0.5612 & 3$_{\Conv}$ & 0 & 0 & {\bf \large +}$_{\Conv}$ & {\bf \large -}            & 0$_{\Conv}$ & 0 & 3 \\      
Gauss        &     & 0.8 & 0.4995 & 0           & 0 & 0 &{\bf \large -}           & {\bf \large -}            & {\bf \large +}           & 0 &   \\
\hline 
asym.L        & 0.1 & 1.0 & 0.5017 & 2$_{\Conv\bullet}$ & 0 & 0 & {\bf \large +}$_{\Conv}$ &  0$_{\Conv}$    & 0$_{\Conv}$  & 0 &  4 (1) \\ 
asym.H        & 1.0 & 1.0 & 0.5144 & 3$_{\Conv}$ & 0 & 0 & {\bf \large +}$_{\Conv}$ &  0$_{\Conv}$    & 0$_{\Conv}$  & 0 &  4 \\
Gauss         &     & 1.0 & 0.4796 & 1           & 0 & 0 & {\bf \large -}           & {\bf \large -}              & {\bf \large +}            & 0 &    \\
\hline
\hline
\multicolumn{4}{l} {\LARGE (D) \fbox{$\lambda=0.6$}} 
\\
\hline
   {\large Pulse}  & {\large $k_E$} & {\large $Q$}  &  {\large $f_{min}(\lambda)$}   & 
            {\large $s_I$} & {\large $s_{h_{CT}}$} & {\large $s_{g_{CC}}$}  & 
            {\large $s_{g_{CB}}$} & {\large $s_{h_{BT}}$}    & 
            {\large $s_{g_{BC}}$} & {\large $s_{g_{BB}}$}    & 
            {\large $d_{H,A-G}$}  \\
\hline
asym.L      & 0.1 & 0.2 & 0.8776 & 2$_{\Conv}$ & {\bf \large -}$_{\Conv}$ & 0 & {\bf \large +} & {\bf \large -}   & 0$_{\Conv}$  & 0 & 3 \\   
asym.H      & 1.0 & 0.2 & 0.8669 & 2$_{\Conv}$ & {\bf \large -}$_{\Conv}$ & 0 & {\bf \large +} & {\bf \large -}   & 0$_{\Conv}$  & 0 & 3 \\
Gauss       &     & 0.2 & 0.8996 & 3           &  0           & 0 & {\bf \large +} & {\bf \large -}   & {\bf \large +}            & 0 &   \\
\hline 
asym.L       & 0.1 & 0.4 & 0.7560 & 2$_{\Conv}$ & 0 & 0 & {\bf \large +} & {\bf \large -} & 0 & 0 & 1 \\
asym.H       & 1.0 & 0.4 & 0.7453 & 2$_{\Conv}$ & 0 & 0 & {\bf \large +} & {\bf \large -} & 0 & 0 & 1 \\
Gauss        &     & 0.4 & 0.8128 & 3           & 0 & 0 & {\bf \large +} & {\bf \large -} & 0 & 0 & \\
\hline 
asym.L       & 0.1 & 0.6 & 0.6609 & 2$_{\Conv}$ & 0 & 0 & {\bf \large +} & 0$_{\Conv}$ & 0 & 0 & 2 \\
asym.H       & 1.0 & 0.6 & 0.6624 & 2$_{\Conv}$ & 0 & 0 & {\bf \large +} & 0$_{\Conv}$ & 0 & 0 & 2 \\
Gauss        &     & 0.6 & 0.7623 & 3           & 0 & 0 & {\bf \large +} &{\bf \large -}           & 0 & 0 & \\
\hline 
asym.L       & 0.1 & 0.8 & 0.5891 & 2$_{\Conv\bullet}$ & 0 & 0 &  {\bf \large +}$_{\Conv}$  & 0$_{\Conv}$  & 0$_{\Conv}$ & 0 & 4 (1)\\
asym.H       & 1.0 & 0.8 & 0.5966 & 3$_{\Conv}$ & 0 & 0 &  {\bf \large +}$_{\Conv}$  & 0$_{\Conv}$  & 0$_{\Conv}$ & 0 & 4 \\
Gauss       &     & 0.8 & 0.5795 & 0           & 0 & 0 & {\bf \large -}              & {\bf \large -}           & {\bf \large +}           & 0 &  \\
\hline 
asym.L       & 0.1 & 1.0 & 0.5350 & 2$_{\Conv\bullet}$ & 0 & 0 & {\bf \large +}$_{\Conv}$ & 0$_{\Conv}$ & 0$_{\Conv}$ & 0 & 4 (1)  \\
asym.H       & 1.0 & 1.0 & 0.5477 & 3$_{\Conv}$ & 0 & 0 & {\bf \large +}$_{\Conv}$ & 0$_{\Conv}$ & 0$_{\Conv}$ & 0 & 4  \\
Gauss  &     & 1.0 & 0.5796 & 1           & 0 & 0 &{\bf \large -}           &{\bf \large -}           & {\bf \large +}           & 0 &    \\
\hline
\hline
\multicolumn{4}{l} {\LARGE (E) \fbox{$\lambda=0.8$}} 
\\ 
\hline
 asym.L     & 0.1 & 0.2 & 0.8922 & 2$_{\Conv}$ & 0 & 0 & {\bf \large +} & {\bf \large -} & 0 & 0 & 1 \\ 
 asym.H     & 1.0 & 0.2 & 0.8814 & 2$_{\Conv}$ & 0 & 0 & {\bf \large +} & {\bf \large -} & 0 & 0 & 1 \\
 Gauss      &     & 0.2 & 0.9148 & 3 & 0 & 0 & {\bf \large +} & {\bf \large -} & 0 & 0 & \\
\hline 
 asym.L       & 0.1 & 0.4 & 0.7701 & 2$_{\Conv}$ & 0 & 0 & {\bf \large +} & 0 & 0 & 0 & 1  \\
 asym.H       & 1.0 & 0.4 & 0.7653 & 2$_{\Conv}$ & 0 & 0 & {\bf \large +} & 0 & 0 & 0 & 1 \\
 Gauss        &     & 0.4 & 0.8394 & 3 & 0 & 0 & {\bf \large +} &{\bf \large -} & 0 & 0 & \\
\hline 
 asym.L     & 0.1 & 0.6 & 0.6809 & 2$_{\Conv}$ & 0 & 0 & {\bf \large +} & 0 & 0 & 0 & 1 \\
 asym.H     & 1.0 & 0.6 & 0.6824 & 2$_{\Conv}$ & 0 & 0 & {\bf \large +} & 0 & 0 & 0 & 1 \\
 Gauss      &     & 0.6 & 0.7926 & 3 & 0 & 0 & {\bf \large +} & 0 & 0 & 0 & \\
\hline  
 asym.L      &  0.1 & 0.8 & 0.6158 & 2$_{\Conv\bullet}$ & 0 & 0 & {\bf \large +}$_{\Conv}$ & 0$_{\Conv}$ & 0$_{\Conv}$ & 0 & 4 (1) \\ 
 asym.H      &  1.0 & 0.8 & 0.6232 & 3$_{\Conv}$ & 0 & 0 & {\bf \large +}$_{\Conv}$ & 0$_{\Conv}$ & 0$_{\Conv}$ & 0 & 4 \\ 
 Gauss       &      & 0.8 & 0.6595 & 0           & 0 & 0 &{\bf \large -} &{\bf \large -} & {\bf \large +} & 0 & \\
\hline
 asym.L    & 0.1 & 1.0 & 0.5684 & 2$_{\Conv\bullet}$ & 0 & 0 & {\bf \large +}$_{\Conv}$ & 0$_{\Conv}$ & 0 & 0 & 3 (1) \\
 asym.H    & 1.0 & 1.0 & 0.5810 & 3$_{\Conv}$ & 0 & 0 & {\bf \large +}$_{\Conv}$ & 0$_{\Conv}$ & 0 & 0 & 3 \\
  Gauss    &     & 1.0 & 0.6796 & 1           & 0 & 0 & {\bf \large -}           &{\bf \large -} & {\bf \large +} & 0 & \\
\hline
\hline
\multicolumn{4}{l} {\LARGE (F)\,\,\, \fbox{$\lambda=1.0$}} 
\\ 
\hline
asym.L  & 0.1 & 0.2 & 0.9000 & 2$_{\Conv}$ & 0 & 0 & {\bf \large +} & $0_{\Conv}$ & 0 & 0 & 2  \\
asym.H  & 1.0 & 0.2 & 0.8922 & 2$_{\Conv}$ & 0 & 0 & {\bf \large +} & $0_{\Conv}$ & 0 & 0 & 2  \\
Gauss   &     & 0.2 & 0.9282 & 3           & 0 & 0 & {\bf \large +} & {\bf \large -}          & 0 & 0 &    \\
\hline
asym.L & 0.1 & 0.4 & 0.7834 & 2$_{\Conv}$ & 0 & 0 & {\bf \large +} & 0 & 0 & 0 & 1 \\
asym.H & 1.0 & 0.4 & 0.7786 & 2$_{\Conv}$ & 0 & 0 & {\bf \large +} & 0 & 0 & 0 & 1 \\
Gauss  &     & 0.4 & 0.8543 & 3           & 0 & 0 & {\bf \large +} & 0 & 0 & 0 &   \\
\hline
asym.L & 0.1 & 0.6 & 0.7009 & 2$_{\Conv}$  & 0 & 0 & {\bf \large +} & 0 & 0 & 0 & 1 \\
asym.H & 1.0 & 0.6 & 0.7024 & 2$_{\Conv}$  & 0 & 0 & {\bf \large +} & 0 & 0 & 0 & 1 \\
Gauss  &     & 0.6 & 0.8126 & 3            & 0 & 0 & {\bf \large +} & 0 & 0 & 0 &   \\
\hline
asym.L & 0.1 & 0.8 & 0.6424 & 2$_{\Conv\bullet}$  & 0 & 0 &  {\bf \large}$_{\Conv}$ & 0$_{\Conv}$ & 0$_{\Conv}$  & 0 & 4 (1) \\
asym.H & 1.0 & 0.8 & 0.6499 & 3$_{\Conv}$         & 0 & 0 &  {\bf \large}$_{\Conv}$ & 0$_{\Conv}$ & 0$_{\Conv}$  & 0 & 4 \\
Gauss  &     & 0.8 & 0.7395 & 0                   & 0 & 0 & {\bf \large -}           & {\bf \large -}          & {\bf \large +}            & 0 & \\
\hline
asym.L & 0.1 & 1.0 & 0.6017 & 2$_{\Conv\bullet}$  & 0 & 0 & {\bf \large +}$_{\Conv}$ & 0$_{\Conv}$  &  0$_{\Conv}$    & 0 & 4 (1) \\
asym.H & 1.0 & 1.0 & 0.6144 & 3$_{\Conv}$         & 0 & 0 & {\bf \large +}$_{\Conv}$ & 0$_{\Conv}$  &  0$_{\Conv}$    & 0 & 4  \\
Gauss  &     & 1.0 & 0.7796 & 1                   & 0 & 0 &{\bf \large -}           &{\bf \large -} &  {\bf \large +} & 0 &    \\
\hline 
\caption{\normalsize Comparison of the optimized results for the asymmetric and Gaussian pulses 
for different parameters of efficiency ($k_E$), regularization ($\lambda$) 
and the strength of therapeutic action ($Q$).  Star symbol is used for the features - components 
of ${\bf s}_{min}$ where the optimal solutions are 
different for symmetrical (line denoted as Gauss) or non-symmetrical 
(line denoted as "asym.") pulses. The rightmost column informs 
about the Hamming distance $d_{H,A-G}$ between the 
optimal configurations corresponding to the asymmetric or 
symmetric pulses, respectively (distance is equal to the total number of the star symbols within 
the line). In addition, the number 
in the parenthesis (included only for nonzero values) 
represents the Hamming distance 
between the asymmetric configurations.} 
\label{tab:asymversG}
\end{longtable}
} 

\subsection{From discrete to continuous optimization on branched manifolds}\label{sec:manifOpt}

In this section we present two illustrative examples of the application 
of the hybrid optimization procedure which demonstrates significance 
of the consecutive corrections to the original
discrete problem. The modification shows that the progress in
the optimization is inseparable from the additional constraints
and parametrizations. 

The original settings given by Eq.(\ref{eq:tildhCT}) can be modified
by means of newly introduced auxiliary six "weight" variables   
{\small
\begin{eqnarray}
\widetilde{h_{CT}}&=& h_{CT} ( 1+ q_{h_{CT}} Q\, h_{CT} s_{h_{CT}} \psi_C)\,, 
\quad 
\widetilde{g_{CC}}= g_{CC} ( 1+  q_{g_{CC}} Q\,  g_{CC} s_{g_{CC}} \psi_C)\,, 
\\
\widetilde{g_{CB}}&=& g_{CB} ( 1+ q_{g_{CB}} Q\, g_{CB} s_{g_{CB}} \psi_C)\,,
\quad 
\widetilde{h_{BT}}= h_{BT} ( 1+  q_{h_{BT}} Q\,  h_{BT} s_{h_{BT}} \psi_B)\,, 
\nonumber 
\\ 
\widetilde{g_{BC}}&=& g_{BC} ( 1+ q_{g_{BC}} Q \, g_{BC} s_{g_{BC}} \psi_B)\,,
\quad 
\widetilde{g_{BB}}=g_{BB} ( 1+ q_{g_{BB}} Q \, g_{BB} s_{g_{BB}} \psi_B)\,.
\nonumber 
\end{eqnarray}}
In order to preserve the original meaning of the unique scalar amplitude $Q$, 
the impact of optimization is redistributed on the non-uniform weights which 
satisfy constraint ${\rm  Tr}({\bf q}) = q_{h_{CT}}+ q_{g_{CC}}+q_{g_{CB}}+q_{h_{BT}}+q_{g_{BC}}+q_{g_{BB}}=6$,
$q_{\bullet}>q_{\mbox{\tiny ground}}>0$.   
The weighting is reflected by the modified penalty term 
$\lambda Q {\bf q}\cdot{\bf s}= \lambda 
Q ( q_{h_{CT}} s_{h_{CT}} $ $ + q_{g_{CC}} s_{g_{CC}} $ $ + 
q_{g_{CB}} s_{g_{CB}} $ $ + q_{h_{BT}} s_{h_{BT}} $ $ + 
q_{g_{BC}} s_{g_{BC}} $ $ + q_{g_{BB}} s_{g_{BB}})$.
The optimization constraints which delimit ${\bf q}$ components 
can be satisfied by considering specific 
parametrization. It can be represented 
by the one-dimensional manifold 
$M_1 \equiv $ $\{{\bf s}\in \Omega,{\bf q}(\eta) \in \mathbb{R}_6,\,
\eta\in (0,1)\subset \mathbb{R} \}$ with the branches  
\begin{eqnarray}
q_{h_{CT}}(\eta)   
&=& q_{\mbox{\tiny ground}}+ (1-q_{\mbox{\tiny ground}}) 
6 \eta^5\,, 
\\
q_{g_{CC}}(\eta)&=&   
q_{\mbox{\tiny ground}}+ (1-q_{\mbox{\tiny ground}}) 30 \eta^4 (1-\eta)\,, 
\nonumber 
\\
q_{g_{CB}}(\eta) &=&  
q_{\mbox{\tiny ground}}+ (1-q_{\mbox{\tiny ground}}) 60 \eta^3 (1-\eta)^2\,, 
\nonumber 
\\
q_{h_{BT}}(\eta) &=& 
q_{\mbox{\tiny ground}}+ (1-q_{\mbox{\tiny ground}}) 60 \eta^2 (1-\eta)^3\,, 
\nonumber 
\\ 
q_{g_{BC}}(\eta) &=& 
q_{\mbox{\tiny ground}}+ 
(1-q_{\mbox{\tiny ground}}) 30 \eta (1-\eta)^4\, 
\nonumber 
\\
\quad q_{g_{BB}}(\eta) &=&
q_{\mbox{\tiny ground}} 
+ (1-q_{\mbox{\tiny ground}}) 6 (1-\eta)^5\,   
\nonumber
\end{eqnarray}
parametrized by the auxiliary real variable $\eta\in \langle 0,1 \rangle$ and constant 
$q_{\mbox{\tiny ground}}\in \langle 0,1\rangle$.  
We note, for later comparisons, that former discrete optimization
corresponding to the uniform weighting $q_{\bullet}=1$ performed 
for the asymmetric pulses and $k_E=1,$ $Q=0.4,$ 
$\lambda=0.2$ leads to ${\bf s}_{min}=[2,-1,0,1,-1,1,0]$ 
corresponding to $f_{min}=0.666259$. Extending the search space on $M_1$ parametrized 
by $q_{\mbox{\tiny ground}}=0.2$, using the Ridders' variant of Dekker-Brent 
method \cite{Ridders1979} applied to dimension $\eta\in \langle 0,1\rangle $ leads 
to the modified ${\bf s}^{M_1}_{min}=[2,-1,1,1,-1,1,0]$ 
but lower $f^{M_1}_{min}=0.5266$ obtained for 
$\eta^{M_1}_{min}\simeq 0.583$. In such case, the highest optimal 
weight $q^{M_1}_{g_{CB},min}\simeq 1.85$ is posed on the OC-OB interaction term
while the lowest weight $q^{M_1}_{g_{BB},min}=0.2605$ 
loses its meaning by being present simultaneously 
with the multiplier $s^{M_1}_{g_{BB},min}=0$.

It should be emphasized that the highest weight is consistent with the central 
role of $s^{M_1}_{g_{CB}}$ in coincidence with the findings of the 
synergistic concept (see sec.\ref{sec:Synerg}). 
As there is no smooth transition to the uniform weighting
for $M_1$, $q_{\mbox{\tiny ground}}<1$, we introduce  
the alternative manifold $M_2$  
by ${\bf q}^{M_2}(\eta) \equiv q_{\mbox{\tiny ground}}[1,1,1,1,1,1] +   
( 1 - q_{\mbox{\tiny ground}}) \, 6 [1,
\eta, \eta^2,\eta^3,\eta^4,\eta^5]/(1+\eta+\eta^2+\eta^3+\eta^4+\eta^5)$ 
to reach $q^{M_2}_{\bullet}=1$ 
in the limit $\eta^{M_2}_{min}=\eta\rightarrow 1$.
However, in this case the optimization routine has determined 
relatively shallow $f^{M_2}_{min}=0.6608$ corresponding to the 
string ${\bf s}^{M_2}_{min}=[2,-1,0,1,-1,1,0]$ 
accompanied by $\eta^{M_2}_{min}=0.86$ and weights   
${\bf q}^{M_2}_{min} \simeq [1.30,1.15,$ $1.03,$ $0.92,0.83,0.75]$.

\subsection{Synergistic quantification of optimization results}\label{sec:Synerg}

The optimization task that we studied clearly demonstrates
that the parameters do not act independently to each other and 
the optimum depends on the simultaneous effect of the parametric combination. 
In the next, we put the multi-parametric optimization outputs into the context
of existing synergistic concepts as conceived in the biomedical and farmacological 
studies of the simultaneous effects of therapeutic drugs 
(see e.g. \cite{Roell2017,Palmer2017}). 
Despite many alternative measures and approaches have been already proposed, 
the synergistic quantification of the optimization outputs 
is not straightforward. The generalization we made 
is inspired by the form of classic Loewe's additivity 
principle \cite{Loewe1926}. 

Before performing quantification of the optimization of the parametric pairs,
we introduce reduced string $\hat{\bf s}$
with the six enumerated components  
$\hat{s}_1=s_{h_{CT}}$, $\hat{s}_2=s_{g_{CC}}$, 
$\hat{s}_3=s_{g_{CB}}$, $\hat{s}_4=s_{h_{BT}}$,
$\hat{s}_5=s_{g_{BC}}$, $\hat{s}_6=s_{g_{BB}}$.
They allow to construct the sets of pairs 
$\hat{\Omega}_{ij}= \{\,\{\hat{s}_i, \hat{s}_j\}$ $;\,\hat{s}_k = 0\, 
\forall k \neq i,\, \forall k\neq j\}$. The extraordinary 
$s_I$ (its zero value does not mean the absence of the pulse influence) 
is excluded from the enumeration, nevertheless, 
its optimization effect is considered as well; see Eq.(\ref{eq:Om4}) below. 
Selection of $\hat{\Omega}_{ij}$ is consistent with the fact   
that {\em pairwise} {\em synergy} makes sense only 
for nonzero pairs $\hat{s}_i$, $\hat{s}_j$. 
The corresponding scalar measures 
may be arranged into $6\times 6$ matrix 
of the generalized {\em  combination} {\em indices} 
\begin{eqnarray}
CIf_{ij}= 
\frac{\hat{f}_{min,\,ij}}{2\hat{f}_{min,\,ii}}  
+\frac{\hat{f}_{min,\,ij}}{2\hat{f}_{min,\,jj}}=
\hat{f}_{min,\,ij}
\underbrace{\left(\frac{2\hat{f}_{min,\,ii} 
\hat{f}_{min,\,jj}}{\hat{f}_{min,\,ii}
+\hat{f}_{min,\,jj}}\right)^{-1}}_{\rm harmonic\,mean}\,,
\label{eq:Sfij}
\end{eqnarray}
where 
\begin{eqnarray}
\hat{f}_{min,\,ij} = 
\min\limits_{\mbox{
$\scriptsize \begin{array}{l}
\{\hat{s}_i,\hat{s}_j\} 
\in \hat{\Omega}_{ij},
s_I\in \Omega^{(4)} 
\end{array}$}}\,\, f(\ldots)\,.
\label{eq:Om4}  
\end{eqnarray}
Here, the single parameter effects are involved in the diagonal elements 
$\hat{f}_{min,\,ii}$, $\hat{f}_{min,\,jj}$, while  
the cumulative effects of the pairs are included in $\hat{f}_{min,ij}$. 
There is an obvious freedom of the choice of the prefactor
[see Eq.(\ref{eq:Sfij})], however, as one may check, 
only the multiplication of both summands by $1/2$ leads
to the harmonic mean normalization of $\hat{f}_{min,ij}$. 

The properties $CIf_{ij}= CIf_{ji}$, $\,\forall i,j$ (symmetry); 
$CIf_{ii}=1$, $\forall i$ (unit diagonal as a referential value corresponding 
to monoparametric effects); $CIf_{ij}>0$ [implicated 
by $f(\ldots)>0$] $CIf_{ij}\leq 1$ $\forall i,j$ can be 
checked easily. At this point, it is also important to 
note the basic difference between the effects of monotherapy 
and the effects caused by the optimization of 
individual parameters. In agreement with the original conception of 
the combination index, lower index value means higher synergy 
of the parametric pair.  
 
\begin{figure}[bth]
\includegraphics[width=0.95\linewidth]{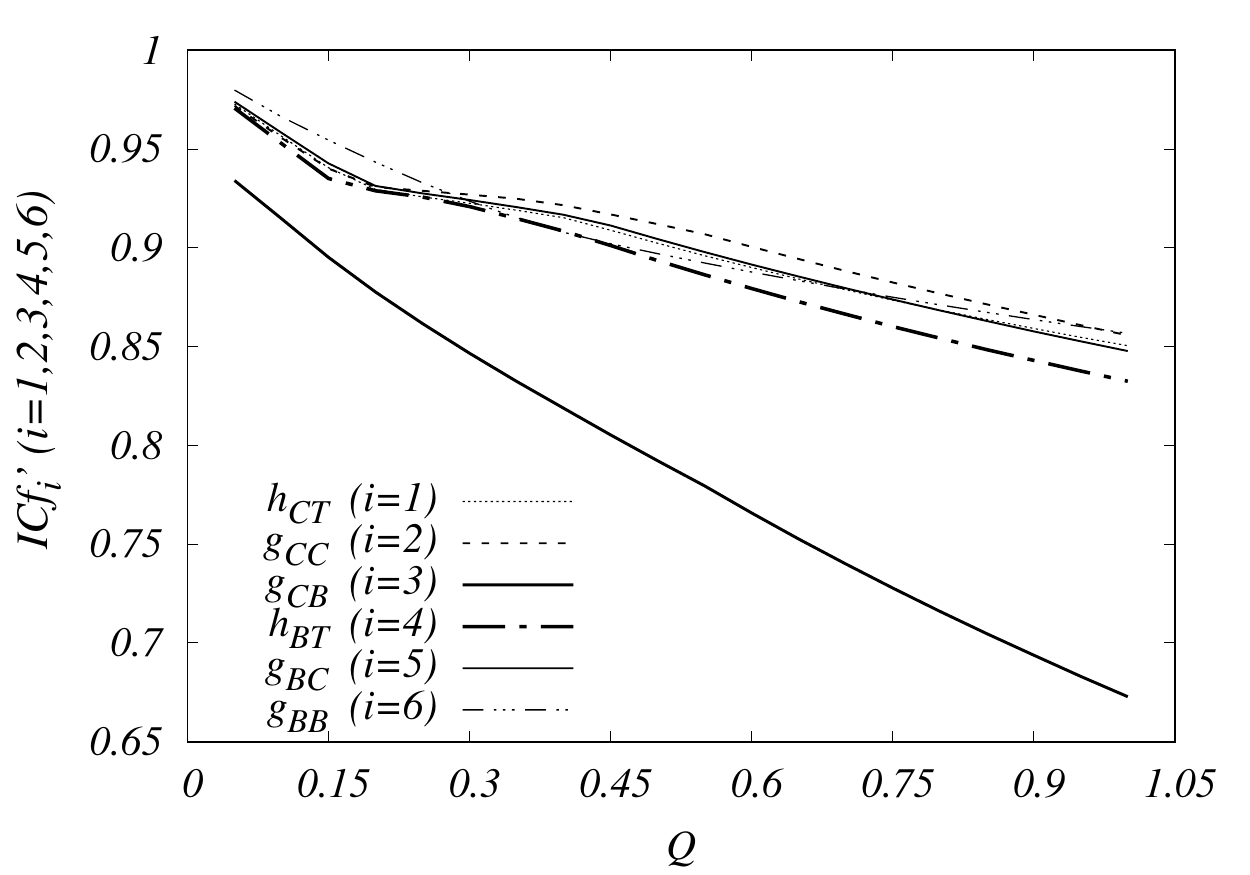}
\caption{The comparison of the integral synergistic 
optimization effects of the parametric 
pairs; $Q$-dependence of $CIf_i'$, $i=1,2,\ldots,6$ 
calculated for the asymmetric pulses characterized by 
$k_E=1$, $W_a=2 day$, $\sigma_a=3 day$, 
$\tau_a=14\, day$.}
\label{fig:SynII}
\end{figure}

To avoid detailed comparison of the pairs, 
we turn to the information comprised in 
$ICf_i'=(1/5)\sum_{j=1; j\neq i}^6 ICf_{ij}$. 
In the correspondence with the standard Loewe's formulation,
the higher the synergy, the lower $CIf_{ij}$ (or $CIf'_{i}$)
is obtained. The results of the numerical calculations of the six 
components of $ICf_i'(Q)$ are depicted in Fig.\ref{fig:SynII}. 
They uncover the highest cumulative synergistic effect of 
$\hat{s}_3=s_{g_{CB}}$ with another parameters. 
The identification of the central role of $s_{g_{CB}}$ in the
optimization process is in the qualitative agreement with 
Table \ref{tab:asymversG}, where the nonzero $s_{g_{CB}}$ persists 
up to the high $\lambda>0$ considered. 

\section{Conceptual model, indirect environmental manipulation}\label{sec:Conc}  

The key aspect of our approach
is the formal separation of the system into the environmental and cancerous 
compartments (populations), as well as the identification of the parameters 
appropriate for indirect environmental therapeutic manipulation.
Specific information contained in the results 
delivered exclusively by simulations 
is hardly transferable between alternative models.
The transmission can be facilitated by the 
generalization obtained by filtering,
comparative analysis and synthesis of some 
particular information contained
in the population models. 

The way how to do it is the conceptualization 
based on general and abstract  
dynamic system for the pair of the population 
vectors $(\mathbfcal{T}(t)$, 
$\mathbfcal{E}(t))$,
where 
$\mathbfcal{T}(t)$ 
describes the tumoral 
and  $\mathbfcal{E}(t)$ 
environmental dynamics, respectively. 
The parametric selection is involved in the constant tuple $\boldsymbol{\chi}_{\mathcal E}$ 
(see Eq.(\ref{eq:listpar}) as particular example). 
Then, the pair of the corresponding rate functions  
$[\mathbfcal{R}_{T}(\mathbfcal{T},\mathbfcal{E}[\boldsymbol{\chi}_{\mathcal E}]),
\mathbfcal{R}_{E}(\mathbfcal{T},\mathbfcal{E}[\boldsymbol{\chi}_{\mathcal E}])]$
characterizes the 
original autonomous 
ODEs system 
\begin{equation}
\frac{d}{dt}
\left(
\begin{array}{l}
\mathbfcal{T}
\\
\mathbfcal{E}
\end{array}
\right) = 
\left(
\begin{array}{l}
\mathbfcal{R}_{\mathcal T}(
\mathbfcal{T},
\mathbfcal{E}[\boldsymbol{\chi}_{\mathcal E}])
\\ 
\mathbfcal{R}_{\mathcal E}(\mathbfcal{T},
\mathbfcal{E}[\boldsymbol{\chi}_{\mathcal E}])
\end{array}\right)\,.
\label{eq:Ne}
\end{equation}
 The exogenous variant of ODEs system belongs to the respective substitutions 
$\boldsymbol{\chi}_{\mathcal E}\rightarrow \widetilde{\boldsymbol{\chi}}_{\mathcal E}(t,{\bf s})$,
$\mathbfcal{E}[
\boldsymbol{\chi}_{\mathcal E}] 
\rightarrow \mathbfcal{E}[
\widetilde{\boldsymbol{\chi}}_{\mathcal E}(t,{\bf s})] $,
$\mathbfcal{R}_X(\mathbfcal{T},\mathbfcal{E}[ \boldsymbol{\chi}_{\mathcal E}]) \rightarrow$
$\widetilde{\mathbfcal{R}}_X(\mathbfcal{T},\mathbfcal{E}[ \widetilde{\boldsymbol{\chi}}_{\mathcal E}(t,{\bf s})])$
for ${\bf s}\in \Omega$ (see Eq.(\ref{eq:partim})) and 
$X \in \{ \mathbfcal{T},\mathbfcal{E}\}$. 
Formally, the solutions 
$[\mathbfcal{T}(t,{\bf s},\mathbfcal{T}_{init},\mathbfcal{E}_{init}) $, 
$\mathbfcal{E}(t,{\bf s},\mathbfcal{T}_{init},\mathbfcal{E}_{init}) ]$
for given initial condition   $ [{\mathbfcal T}(t_{init}), $ 
${\mathbfcal E}(t_{init})] \equiv$ 
$  [{\mathbfcal T}_{init},{\mathbfcal E}_{init}]$ with $t_{init} \leq t_S$
can be obtained. 
Subsequently, the adverse effects 
of the tumoral dynamics
along the path segment 
{\small
\begin{equation}
{\mathcal{P}}_{\mathbfcal{T},\mathbfcal{E}}(t_S,t_E,{\bf s})
\equiv  \left\{\, \left[\mathbfcal{T}(t,{\bf s},\mathbfcal{T}_{init},\mathbfcal{E}_{init}), 
\mathbfcal{E}(t,{\bf s},\mathbfcal{T}_{init},\mathbfcal{E}_{init})\right]
 \,,\, \forall t\in \langle  t_S, t_E \rangle\,\right\}
\end{equation}}  
are quantifiable 
by the objective function 
\begin{equation}
f(t_S,t_E,{\bf s} ) = {\mathcal F}\{\mathcal{P}_{\mathbfcal{T},\mathbfcal{E}}(t_S,t_E,{\bf s}) \}\,,
\end{equation}
where ${\mathcal F}\{.\}$ stands for some real-valued functional. 
In particular cases, that we have already discussed (see Eq.(\ref{eq:fTJ})), 
the special emphasis has been placed on the boundary interval values.
Again, as in the case of specific formulation in sec.\ref{sec:methodology}, 
the optimal therapy ${\bf s}_{min}$
is determined by Eq.(\ref{eq:smin}). 

\section{Conclusion} 
 
In this article we present a global strategy that combines sensitivity analysis 
and global optimization approach which contrasts with the single parameter (local) investigations. 
As an intriguing aspect of our holistic view we see the synthesis of several different research directions.
Substantial nonlinearity of ODE system is studied using multi-parametric 
pulses of symmetric and asymmetric form. In general terms, our approach  
provides a mapping between continuous problem formulated as ODEs 
and discrete problem where the ordering of the solutions according to the 
values of the respective objective function matters. Apart from the optimal 
therapeutic procedure itself, we are interested in the analysis of the sensitivity
of differential equation systems to the complex parametric changes to which 
the potential therapies can be targeted. The results support potential 
benefits of our approach not only for monotherapies but also for the combined treatments. 
By incorporating synergistic measures of the parametric pairs, progress has been made  
in the model-level understanding of the complex systemic responses to the multi-dimensional pulses.
It seems that many models of biological systems could be studied in a similar way 
regardless of the area of application. From a biological/pharmacological point 
of view, we do not follow conventional line of direct elimination of cancer cells; 
instead, more emphasis is given to the indirect effect achieved by 
the optimized manipulation with the tumoral environment. In addition, we point 
out that the results of our discrete combinatorial approach reveal 
many possible alternatives of restoring the balance between myeloma and the 
environment, as opposed to achieving higher optimality for a narrow set 
of specialized therapies.

Biological relevance of here analyzed MM system derives from the models by 
Komarova et al. \cite{Komarova2003} and Koenders and Saso \cite{Koenders2016}.
This model for studying tumor - bone interactions is built in accordance with the general parsimony 
framework \cite{Pivonka2010} which captures the fundamentals of the system. 
Presented approaches contribute to the understanding of the respective nonlinear ODEs
by analyzing their complex responses. The works which focus on the problem of cancer treatment and its modeling can be roughly divided along two principal
therapeutic targets: the tumor or its environment, respectively \cite{Ayati2010}.
Our present work is in line with environmental aspect of cancer progression which is 
profoundly stressed in evolutionary and ecological models of cancer \cite{Merlo2006,Anderson2006,Greaves2007,Gatenby2009a}.
We hope that the methodology presented in this work will be extended to take into account intratumor heterogeneity \cite{Wodarz2005,Amor2014,Horvath2016}, 
e.g. cell-to-cell heterogeneity at the MM level \cite{Paino2014,Bouchnita2016,Tang2016}. 
The indirect, environmentally oriented therapies can cause the erosion of the habitats of the cancer clones, or,
eventually, to decrease the probability of drug resistance. 
Many promising and innovated therapies exploit indirect 
therapeutic effects, such as the cell cycle manipulation,
influencing or remodeling of vascular networks \cite{Swierniak2009,Byrne2010}.
More specifically, in Ref. \cite{Modi2012} the authors identified 
indirect antimyeloma mechanisms preventing bone resorption \cite{Modi2012}. 
Another example which should be mentioned in this context is immunotherapy \cite{Cappucio2007,Castiglione2006,Pillis2006,Reisfeld2013}, 
where particular emphasis is placed on tumor-associated macro\-phages and cancer-associated fibroblasts 
as the determinants of the tumor microenvironment. The recent literature \cite{Terpos2018} reminds that homeostasis of normal 
bone tissue formed by OBs, OCs and osteocytes is regulated 
predominantly by osteocyte cells derived from OBs. 
In the case of MM, homeostasis is impaired, leading to the bone resorption, 
pathogenic lesions and tumor expansion. We assume that this type of knowledge  
can be appropriately integrated into the three-component form of the environmental 
population with eventual therapeutic influence on tumorigenesis.

\section*{Acknowledgments}
This work was supported by (a)~APVV-15-0485 by Slovak Research and Development Agency;(b)~VEGA  No. 1/0250/18; (c) VEGA 1/0156/18 
by Scientific Grant Agency of the Ministry of Education of Slovak Republic. 

\appendix
\section{Appendices}
\subsection{Robustness of the optimum, reoptimization in the periodic environment}\label{Ap:Period}

The therapies can be delivered in periodic or other appropriate manner \cite{Foo2009,Ananthakrishnan2010} 
instead of in one time only. As an alternative to focusing on the optimal number of the pulse repeats, 
we investigate the feasibility and robustness of the periodic extension of the results 
achieved for single pulse [see sec.\ref{sub:sympRes}] without reoptimization 
for the periodic conditions. The extension represents a kind of myopic, 
short-term strategy \cite{Handa2008}. 
 
Technically, the extension we applied is based on ${\bf s}_{min}$ and period $t_E-t_S$. 
The transient, not purely periodic, $C(t), B(t), T(t),J(t)$ responses are obtained when the 
integration of Eqs.(\ref{eq:CBTJ}), (\ref{eq:RCRBRTRJ}) is performed for the 
sequence of the joint intervals $\langle t_0^{(per)}(m)$, $t_1^{(per)}(m)\rangle$, 
where $ m = 0, 1, 2, \ldots $ are indices of the periodic shifts $ t_0^{(per)}(m) = t_0 + m ( t_E-t_S)$, 
$t_1^{(per)}(m)=t_1 +  m ( t_E-t_S) $ in the locations of the peaks replacing $t_0$, $t_1$ in the functions 
$\psi(t,t_0\rightarrow t_0^{(per)}(m),\ldots)$ and $\psi(t,t_1\rightarrow t_1^{(per)}(m),\ldots)$. 

\begin{figure}[bth]
\includegraphics[width=0.8\linewidth]{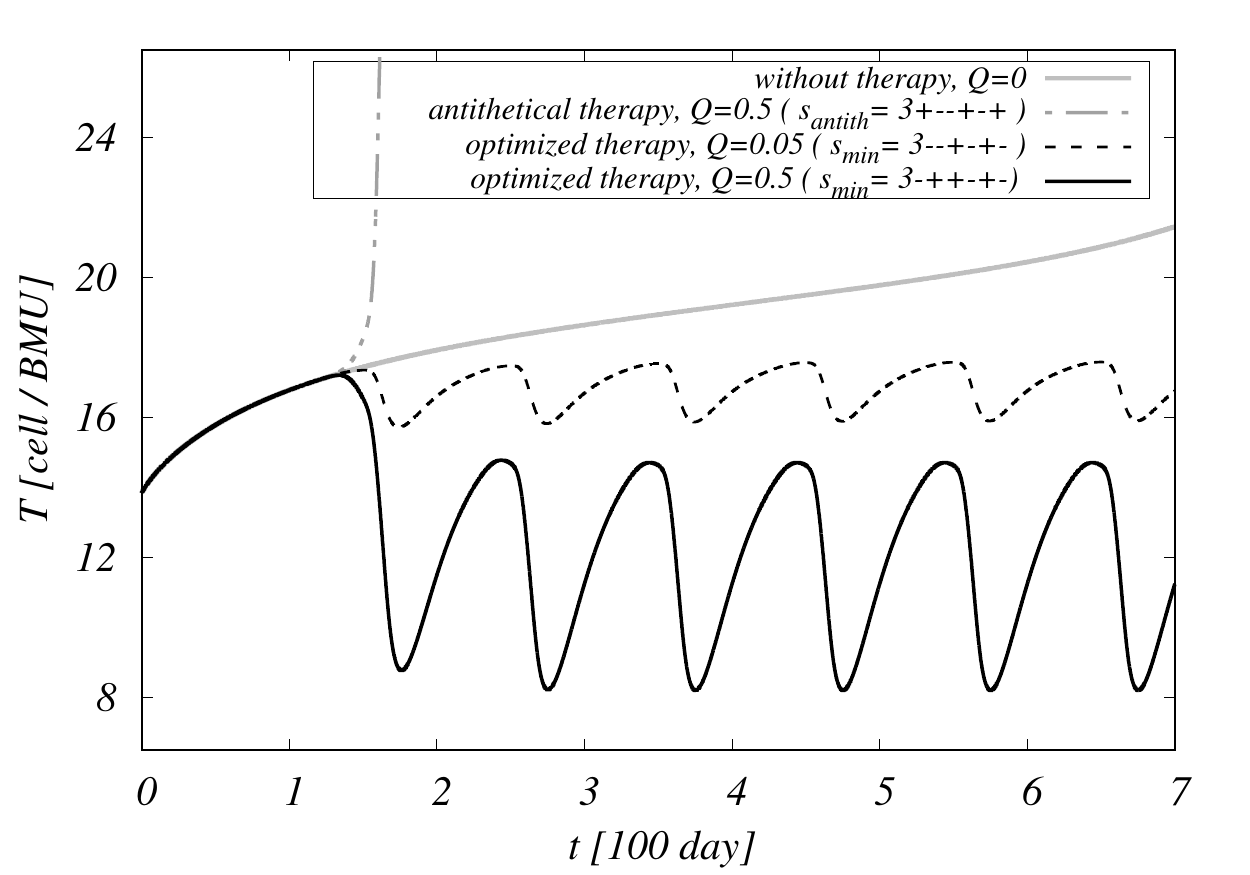}
\caption{The illustrative treatment of the tumor  
by means of the periodic environmental changes. 
The comparison for several $Q$ values. 
The periodic extension of the solution obtained for 
$Q=0.5$ is based on the optimal string 
${\bf s}_{min}=$ $ [ 3, -1,+1$  $,+1,-1,+1,-1]$ 
(optimized for the non-periodic case only). 
The "antithetical" virtual therapy 
${\bf s}_{antith}=
[ 3,+1,-1,$ $-1,+1,-1,+1]$ (where component  $\mp 1$ is replaced by $\pm 1$) has been chosen for the comparative purposes. 
Note that this therapy is not necessarily 
the worst possible therapy obtained for $Q=0.5$. 
The neutral [no influence 
of $\Psi_B(.)$ or $\Psi_C(.)$ terms] 
choice calculated for $Q=0$ corresponds to the 
tumor growth without environmental moves.} 
\label{fig:ff7}
\end{figure}

\begin{figure}[bth]
\includegraphics[width=\linewidth]{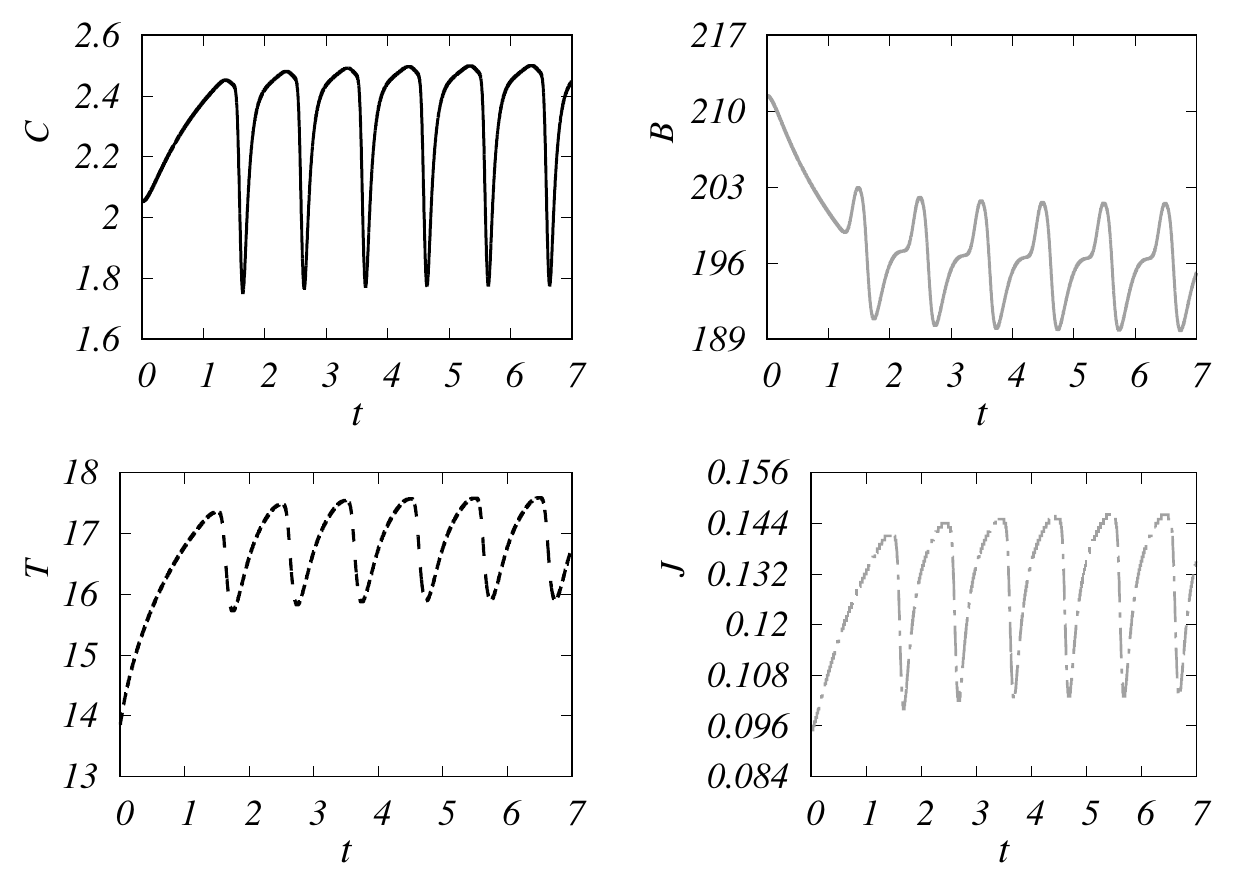}
\caption{The onset of the periodic dynamics of the four cell populations 
characterized by $C, B$, $T, J$ (population numbers per BMU) induced
by the exogenous periodic stimuli with 
the period $t_{E}-t_{S}$. Calculated for $Q=0.05$ 
and the respective string ${\bf s}_{min}=[3,-1,-1,+1,-1,+1,-1]$.} 
\label{fig:ff8}
\end{figure}

From Figs.(\ref{fig:ff7}) and (\ref{fig:ff8}) we see that despite 
the optimization has not been performed in the 
periodic case and only suboptimal ${\bf s}_{min}$ has been used,  it turns out that the therapeutic interventions obtained 
are sufficient to keep tumors within safe limits.

\subsection{Stability of integrators}\label{Ap:Stab} 

\begin{figure}[bth]
\includegraphics[width=0.95\linewidth]{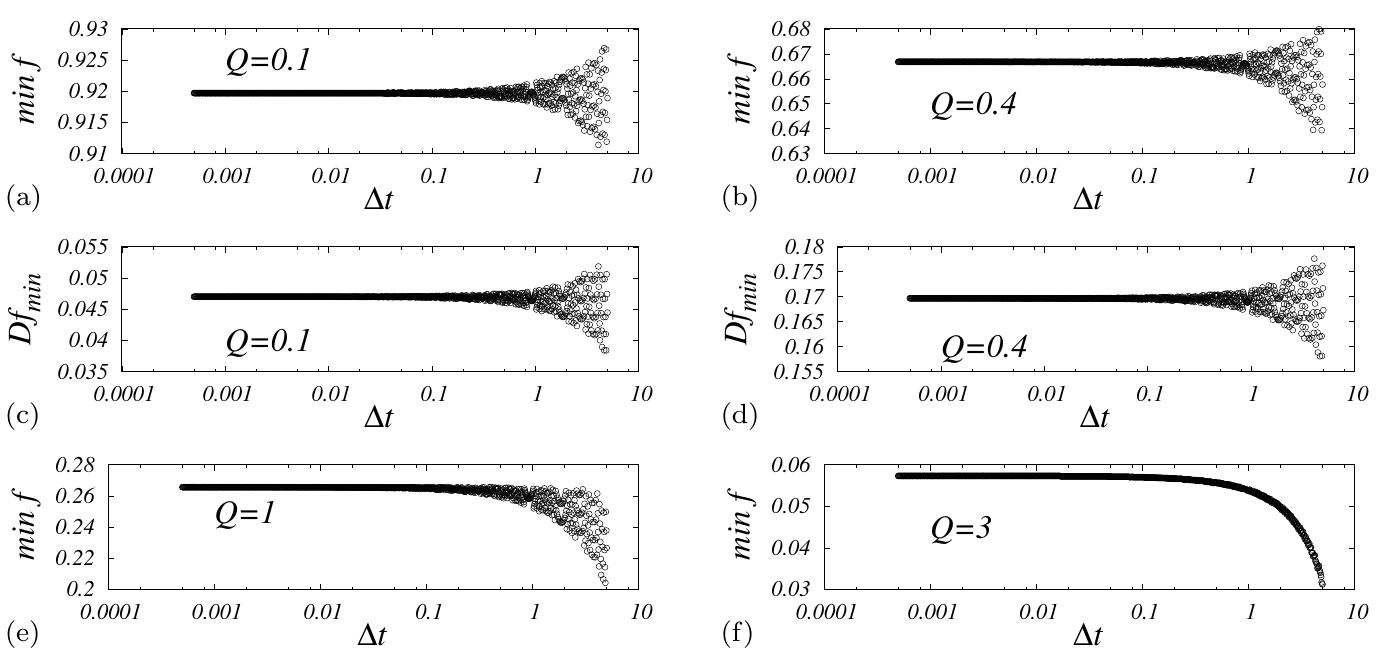}
\caption{This figure shows results of the combined testing 
with the typical loss of numerical stability corresponding
to the optimized outputs of the RK4 method obtained by the gradually
increasing time integration step. Calculated for $n_0=0$ 
and $Q\in \{0.1,0.4,1,3\}$. Data in panels (a), (b), (e), (f) show $\min f$ while (c),(d) 
panels show $Df_{min}$ quantity. As it is obvious from the panel (f), the instability may
also have a one-sided character.
Note that dependencies (a) - (e) look seemingly stochastic for 
$ \Delta t \gtrapprox 0.1$, but detailed view reveals certain regularity of patterns 
in this region.}
\label{fig:ffA}
\end{figure}

In the numerical ODE analysis, the explicit integration methods assume that
almost any numerical precision can be achieved by selecting 
small enough integration step. Therefore, if the explicit methods,
such as the above mentioned RK4 are applied, great care 
must be taken to guarantee the convergence and the asymptotic
properties of iterates, as these may exhibit strong dependence on the integration
step size.   More specifically, the
problem known as ODE stiffness occurs in the cases where approximate errors
of the numerical method cannot be  sufficiently
suppressed regardless of how small the integration step is.
Therefore, robust implicit integrators need to be used to verify solutions 
obtained using explicit methods. For the ODE application 
described above we perform combined
problem-specific testing using the overall integration-optimization
procedure instead of purely numerical integration without locating
optima. The aim of the combined testing [see Fig.\ref{fig:ffA}]
is to examine the persistence of $\min f$ (and the related
metrics $Df_{min}$), as well as revealing potential 
stiffness artifacts.

In Table \ref{tab:tabA} we list the thresholds in the parameter $\Delta t$
which indicate the onset of instabilities of the respective 
integrators. The numerical findings for RK4 are provided along with Euler
predictor-corrector (PEC) and PECECE 
variants with recurrent corrector steps.  Both alternative integrators 
can be classified 
as implicit due to the incorporation 
of the trapezoidal rule. 
By comparing the tabulated
thresholds obtained 
for several $ Q $ we have obtained a universal 
lower estimate of $ \Delta t \sim \mathcal {O} ($ 0.01). 
It is obvious that the use of this or smaller step size 
does not pose the risk of substantial stiffness errors 
for any of the above methods, which agree in $ {\bf s}_ {min}$.
This means that our original choice of 
$ \Delta t = 5 \times 10 ^ {- 4} $ for RK4 
provided reliable results.

 \begin{table}[bth]
\begin{center}
\begin{tabular}{|l| |m{1.3cm} m{1.7cm} | m{1.3cm} m{1.7cm} | m{1.4cm} m{1.75cm} | } 
\hline           
\vspace*{-2.2mm}
$Q$   & RK4 &  RK4 &  PEC  &  PEC  & PECECE &   PECECE 
\\
& \mbox{\small $rtol$}   &   \mbox{\small $rtol$} 
&  \mbox{\small $rtol$}  &   \mbox{\small $rtol$}
&  \mbox{\small $rtol$}  &   \mbox{\small $rtol$} \\
\vspace*{-6mm} \\
& \mbox{\small $=10^{-3}$}   &   \mbox{\small $=2\times 10^{-3}$} 
&  \mbox{\small $=10^{-3}$}  &   \mbox{\small $=2\times 10^{-3}$}
&  \mbox{\small $=10^{-3}$}  &   \mbox{\small $=2\times 10^{-3}$} \\
\hline
0.1 &  0.496 &  0.966  &   0.530  &  1.036  &  0.506    &  1.025 \\
0.4 &  0.116  & 0.219  &  0.146   &  0.279  &  0.146    &  0.291 \\
1.0 &  0.024  & 0.045  &  0.042   &  0.077  &  0.042    &  0.077 \\
3.0 &  0.022  & 0.038  &  0.158  &   0.248   & 0.276    &  0.303 \\
\hline
\end{tabular}
\end{center}
\caption{\label{tab:tabA} The table contains list of the threshold 
values of the parameter $\Delta t$ defined as the lowest values for which the 
relative change of $\min f$ reaches at least the predefined levels of relative tolerance 
$rtol= 10^{-3}$ (or $2\times 10^{-3}$) with respect
to the reference (saturated) value of $\min f$ obtained for sufficiently small $\Delta t = 5\times 10^{-4}$.
We see that PEC stability exceeds those of RK4, moreover, adding another iteration
loop within PECECE seems to be even redundant for given parameters
and intervals studied. The advantage of implicit methods is 
more obvious at $Q=1$, where the instabilities are stronger 
(see Fig.\ref{fig:ff6}).  The calculations are performed for $n_0=0$.}
\end{table}

\pagebreak

\vspace*{1mm}

\end{document}